\documentclass[journal=jctcce,manuscript=article]{achemso}

\usepackage{amsmath}
\usepackage{amssymb}
\usepackage{graphicx}

\usepackage{dcolumn}
\usepackage{bm}
\usepackage{longtable}
\usepackage[usenames, dvipsnames]{color}
\usepackage{subfigure}
\usepackage{multirow}
\usepackage{dsfont}
\usepackage{ulem}
\usepackage{xcolor}

\author{Ivan Duchemin}
\email{ivan.duchemin@cea.fr}
\affiliation{Univ. Grenoble Alpes, CEA, IRIG-MEM-L\_Sim, 38054 Grenoble, France}
\author{Xavier Blase}
\affiliation{Univ. Grenoble Alpes, CNRS, Inst NEEL, F-38042 Grenoble, France}

\title{ Robust  analytic continuation approach to many-body $GW$ calculations}

\date{\today}
\begin{document}

\begin{abstract}
The analytic continuation of the $GW$ self-energy from the imaginary to the real energy axis is a central difficulty for approaches exploiting  the favourable properties  of response functions at imaginary frequencies. Within a scheme merging contour-deformation and analytic-continuation techniques, we show on the basis of extensive calculations for large molecular sets that it is preferable to perform  an analytic continuation of the dynamically screened Coulomb potential $W$  rather than the much more structured self-energy operator. The case of states lying far away from the gap, including core states, is addressed by generalizing the analytic continuation scheme, accounting further for quasiparticle lifetimes. 
 \end{abstract}



\section{Introduction}

The $GW$ approximation \cite{Hed65,Str80,Hyb86,God88,Farid,Ary98,Oni02,ReiningBook} to the exchange-correlation self-energy stands nowadays as a very popular formalism to calculate charged excitations,  as obtained e.g. using photoemission techniques,  in  metallic or semiconducting, finite size or periodic   systems. Belonging to the family of Green's function many-body perturbation theories (MBPT), the related formalism dates back to the mid-60s  with a pioneering application to the electronic properties of the interacting homogeneous  electron gas.  \cite{Hed65}
Nowadays, efficient implementations, including algorithms with cubic \cite{Foe11,Liu16,Wil18}  or sub-cubic \cite{Neu14}   scaling with system size, paved the way to applications to  large   systems with up to several hundred atoms. \cite{Neu14,Gov15,Li16,Gao16,Wil18} Besides providing accurate electronic energy levels, the $GW$ formalism has been used further to evaluate the quasiparticle lifetime with respect to electron-electron scattering \cite{Cam99,Key00,Mar02,Yi10}  or provide more accurate electron-phonon coupling matrix elements as compared to DFT calculations performed with standard semilocal functionals. \cite{Laz08,Fab11,Yin13,Li19}

At the core of  Green's function MBPT stands the self-energy operator $\Sigma({\bf r},{\bf r}';E)$ that accounts for exchange and correlation interactions beyond the classical Hartree term. An important feature of such an operator, as compared e.g. to the DFT exchange correlation or the Fock exact exchange potentials, lies in its energy dependence. In short, the self-energy operator must be taken at the energy of the state it is acting on. Considering the
specific case of the $GW$ approximation, the self-energy reads:
 
\begin{equation}
\Sigma({\bf r},{\bf r}' ; E) = \frac{i}{2\pi} \int d\omega \;  e^{i  \eta \omega  } G({\bf r},{\bf r}' ; E+ \omega) W({\bf r},{\bf r}' ;  \omega) 
\label{sigma0}
\end{equation}
where $W$ is the time-ordered screened-Coulomb potential and $G$ the time-ordered single-particle Green's function 
($\eta$ is a small positive infinitesimal). The treatment of the energy integration involved in the construction of the $GW$ self-energy operator stands as a central difficulty as it requires, in principle, the knowledge of the dynamically screened Coulomb potential along the real frequency axis where $G$ and $W$ present a large number of poles. Such an integral can be performed analytically provided that one knows the exact spectral representation of the screened Coulomb operator. This representation can be obtained by calculating explicitly  the poles and eigenvectors of the susceptibility operator, within e.g. the random phase approximation (RPA), at the cost of diagonalizing a matrix of which the size grows quadratically with system size,   namely a formally   $O(N^6)$ process. \cite{Tia06,Set13,Bru16,Ver18}


As a simplified but related  technique, plasmon-pole approximations \cite{Hyb86,Lin88,God89,Eng93,Bla14,Roh95} provide a much simplified explicit spectral representation that has been used in the early days of the $GW$ formalism. \cite{Hyb86,God89} While plasmon-pole approximations are extremely efficient numerically,  their accuracy can be challenged in systems where response functions are not clearly dominated by well defined plasmon peaks. This is the case e.g.  in small molecular systems where discrete electron-hole excitations may not merge into broad collective modes.

Numerical integrations by explicitly calculating the susceptibility on discretized frequency grids provide  alternative solutions.  While integration along the real axis has been implemented in a few codes, \cite{Miy00,Shi06,Ros10,Liu15}
 calculation of $W$ along the imaginary axis  - where all operators present a much smoother frequency dependence - has been also proposed in the very early days
of   $GW$ calculations. \cite{God88} Further, the so-called space-time technique, \cite{Roj95,Uma09,Wil18}  with its recent extensions in quantum chemistry within the framework of interpolative density fitting or separable resolution-of-the-identity approaches, \cite{Lu15,Lu17,Duc19} leads to the calculation of the susceptibility at imaginary times. Using Laplace transform,  the susceptibility and related screened Coulomb potential can be then obtained along the imaginary-frequency axis. 

From the knowledge of the screened Coulomb potential along the imaginary axis, one may calculate the self-energy  operator at imaginary energies and proceed with its analytic continuation (AC) to the real axis where it is needed. 
\cite{Roj95,fhiac,Uma10,Ngu12,Pha13,Chu16,Nab16,Wil18} 
Such an approach is extremely elegant and numerically efficient, but difficulties exist concerning its accuracy in relation with the pole structure of the $GW$ operator at real frequencies.\cite{Can16} In a recent benchmark study of small molecular systems, \cite{Set15} the need to increase the number of Pad{\'e} approximants to a large number was documented for several systems. Even with more than a hundred approximants, pathological systems could not be treated with the desired accuracy  in the most simple case of frontier orbitals.  

Finally, in the contour deformation technique, \cite{God88,Far88,Leb03,Bla11,Gov15,Gol18} the integral along the real-axis is transformed  into an integral along the imaginary axis. This integral must be  complemented however by the contribution of the residues of $W$ at the poles of the Green's function $G$ that are shifted into the first/third quadrant by the energy at which the self-energy is required (see Fig.~\ref{fig1}). The contour-deformation approach requires thus the knowledge of the screened Coulomb potential along both the imaginary and the real axis.

In the present study, we explore an approach merging the contour deformation technique with an analytic continuation of the screened Coulomb potential $W$ to the complex plane for the evaluation of the needed residues. This  leads to a very robust scheme as compared to the analytic continuation of the much more structured self-energy $\Sigma$.  Our approach is validated by exploring its performances over the $GW$100 test set~\cite{Set15,Kra15,Car16,Mag17,Gov18} and a recent set of medium size molecules, \cite{Kni16} including  systems that were shown to be pathological for the direct analytic continuation of the self-energy.  We show that very few calculations of the susceptibility along the imaginary axis are required  to reach the meV accuracy in the calculation of the quasiparticle energy of states lying close to the gap,  in particular the highest occupied (HOMO) and lowest unoccupied (LUMO) molecular orbitals. Further, the $GW$ self-energy and associated spectral functions can be efficiently and accurately obtained over a large energy range, covering e.g. the full valence manifold,  provided that (a) a few additional reference $W(\omega)$ matrices are calculated over a coarse energy grid in the complex plane with $\Re(\omega)$ spanning the targeted energy window  and (b) the lifetime of quasiparticle states is accounted for.  This approach is finally adapted to the case of core levels. The present study extends to the $GW$ formalism applied to large sets of molecular systems the recent scheme proposed in the $GT$ study of solid iron, \cite{Fri19} generalizing further the hybrid contour-deformation/analytic-continuation scheme to deep lying states. 


\section{Theory}

We briefly recall here below the main characteristics of the general $GW$ formalism as well as the contour deformation approach that we combine with an analytic continuation (AC) of the screened Coulomb potential. Details of the integration grid along the imaginary axis are provided, together with the technical details associated with the calculations presented in the Section ``Results".

\subsection{The $GW$ formalism}



In a standard $GW$ calculation, one starts from an input time-ordered one-body Green's function typically built from  $\lbrace \varepsilon_n, \phi_n \rbrace$  Kohn-Sham eigenstates, namely: 
 \begin{eqnarray}
 G({\bf r},{\bf r}'; \omega ) =  \sum_{n}    \frac{   \phi_n({\bf r})  \phi_n^*( {\bf r}')  }{ {\omega} - \varepsilon_n +i \eta \times \text{sgn}(\varepsilon_n -E_F) }  
 \label{GreenFunc}
 \end{eqnarray}
 where $\eta$ is a small positive infinitesimal and $E_F$ the Fermi energy. The corresponding $GW$ self-energy $\Sigma({\bf r},{\bf r}' ; E)$ is then generally divided into exact exchange $\Sigma^X({\bf r},{\bf r}' ; E)$ and correlation-only $\Sigma^C({\bf r},{\bf r}' ; E)$ contributions.
 The exact exchange operator built from the one-body eigenstates can be formulated as:
 \begin{equation}
\Sigma^X({\bf r},{\bf r}' ; E) = \frac{i}{2 \pi} \int_{-\infty}^{+\infty} d\omega \;  e^{i  \eta \omega  } G({\bf r},{\bf r}' ; E+ \omega) V({\bf r},{\bf r}') 
\label{exx}
\end{equation}
 with $V$ the bare Coulomb potential, so that by writing $\widetilde{W}=W-V$, the correlation-only self-energy operator becomes:
 \begin{equation}
\Sigma^C({\bf r},{\bf r}' ; E) = \frac{i}{2 \pi} \int_{-\infty}^{+\infty} d\omega \;  e^{i  \eta \omega  } G({\bf r},{\bf r}' ; E+ \omega) \widetilde{W}({\bf r},{\bf r}' ;  \omega)
\label{sigmaC}
\end{equation}

In order to highlight the desirable features  of the present approach, one can take on general grounds the (generally unknown) spectral representation of $\widetilde{W}$:
 \begin{eqnarray}
\widetilde{W}({\bf r},{\bf r}'; \omega ) = \sum_{\lambda}   \frac{  w_{\lambda}({\bf r})  w^*_{\lambda}({\bf r}')  }{ \omega - \Omega_{\lambda} + i \eta } - \frac{ w_{\lambda}({\bf r})  w^*_{\lambda}({\bf r}') }{ \omega + \Omega_{\lambda} - i \eta }  
\label{WFunc}
\end{eqnarray}
with $\lbrace \Omega_{\lambda}, w_{\lambda}({\bf r}) \rbrace$  the related (positive) pole energies and amplitudes, indexed by $\lambda$. Together with the spectral representation of $G$, the knowledge of the spectral representation of $\widetilde{W}$ allows the analytical integration of Eq.~\ref{sigmaC}, yielding for the expression of $\Sigma^C({\bf r},{\bf r}' ; E)$:
 \begin{equation}
 \begin{split}
 \Sigma^C({\bf r},{\bf r}' ; E) = & \sum_{i,\lambda} \frac{ \phi_i({\bf r})w_{\lambda}({\bf r})  w^*_{\lambda}({\bf r}')\phi^*_i({\bf r}')  }{  E - \varepsilon_i + \Omega_{\lambda}  -i \eta  }  \\
     +& \sum_{a,\lambda} \frac{ \phi_a({\bf r})w_{\lambda}({\bf r})  w^*_{\lambda}({\bf r}')\phi^*_a({\bf r}') }{ E - \varepsilon_a - \Omega_{\lambda}  + i \eta  } 
\end{split}
\end{equation}
where i/a index occupied/unoccupied input (e.g. Kohn-Sham) eigenstates. As such, the resulting $\Sigma^C({\bf r},{\bf r}' ; E)$ exhibits numerous poles along the real axis, distributed at energies $\lbrace \varepsilon_i - \Omega_{\lambda} \rbrace$ and $\lbrace \varepsilon_a + \Omega_{\lambda} \rbrace$, summing up to the number of poles of $G$ times the number of poles of $W$. 
Such a large number of poles may lead  to potential difficulties with the standard AC approach that attempts to obtain $\Sigma^C({\bf r},{\bf r}' ; E)$ from the $\Sigma^C({\bf r},{\bf r}' ; i\omega)$ calculated along the imaginary frequency axis.

\subsection{The contour-deformation approach}

\begin{figure}[h]
\includegraphics[width=8cm]{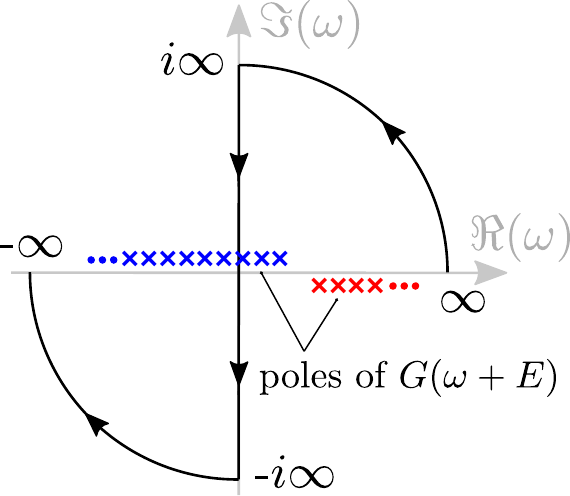}
\caption{Contour-deformation technique. The  poles of the  $G(E+\omega)$  time-ordered Green's function for occupied/unoccupied states are indicated in blue/red.  When the energy $E$ is smaller than the HOMO energy, which happens when we seek the quasiparticle energy for occupied states,  poles of $G(E+\omega)$ start entering the first quadrant (case of the figure).}     
\label{fig1}
\end{figure}

Contrary to the input Green's function $G$, the spectral representation of $\widetilde{W}$ is generally unknown and one has to perform integration of Eq.~\ref{sigmaC} numerically. As a way to proceed, 
the contour-deformation approach transforms this latter integral along the real-frequency axis as an integral along the imaginary-frequency axis plus residues associated with the poles of $G(E+\omega)$ that have entered the first and third quadrant (see Fig.\ref{fig1}), namely:
\begin{equation}
\begin{split}
\Sigma^C({\bf r},{\bf r}'; E ) = - & \frac{1}{2 \pi}\int_{-\infty}^{+\infty} d\omega \; G({\bf r},{\bf r}' ; E+ i\omega) \widetilde{W}({\bf r},{\bf r}' ;  i\omega) \\ 
- & \sum_i \phi_i({\bf r}) \phi_i^*({\bf r}') \widetilde{W}({\bf r},{\bf r}'; \varepsilon_i - E ) \theta(\varepsilon_i-E) \\
+ & \sum_a \phi_a({\bf r}) \phi_a^*({\bf r}') \widetilde{W}({\bf r},{\bf r}'; E - \varepsilon_a ) \theta(E-\varepsilon_a) 
\end{split}
\label{anacon0}
\end{equation}
with $\theta$ the  Heaviside function and $\varepsilon_i / \varepsilon_a$ the occupied/unoccupied  eigenstates used to build $G$. This reformulation is exact, leading to accurate results provided that the numerical treatment of the frequency integral is converged. In fact, excellent accuracy can be achieved  with few quadrature points due to the absence of poles of $\widetilde W$ along the imaginary axis.  

The main drawback of the contour-deformation approach is that the explicit calculation of the required $\widetilde{W}({\bf r},{\bf r}'; \varepsilon_n - E )$ residues can become a bottleneck if the energy $E$ at which the self-energy is required lands outside of the HOMO-LUMO gap, yielding potentially a large number of poles of $G(E+\omega)$ in the first/third quadrant.
Indeed, the number of Kohn-Sham eigenstates within a given energy window  grows linearly with system size, and so does the number of associated residues. Assuming a standard calculation of the independent-electron operator $\chi_0({\bf r},{\bf r}'; \omega)$ that grows as $\mathcal{O}(N^4)$ with system size, the total cost of calculating these residues grows thus as $\mathcal{O}(N^5)$.
In addition, for a given system, as the energy window associated with the two Heaviside functions increases for energies  located far away from the gap, the  number of residues to be considered grows. 
Namely, the prefactor associated with the $\mathcal{O}(N^5)$ scaling increases as the states for which $\Sigma^C$ is calculated go deeper into the occupied or unoccupied manifold.  

To bypass these difficulties, we explore an approach where the residues $\widetilde{W}({\bf r},{\bf r}'; \varepsilon_n - E )$\ are obtained by using an analytic continuation (AC) of the screened Coulomb potential $W$ to the real-energy axis. This approach has several advantages: i) as argued and demonstrated in the previous section, the screened Coulomb potential is much less structured than the self-energy, considerably stabilising the AC procedure; ii) it can be done with no additional cost when continuation is performed from the imaginary axis only 
since the integration in Eq.~\ref{anacon0} already requires the explicit calculation of $W(i\omega)$ for a set of imaginary frequencies (Fig.~\ref{fig_W_sampling}a); iii) contrary to the case of $\Sigma_C(\omega)$, the computational cost of $W(\omega)$ is constant all over the complex plane, so that extra reference points can be added away from the imaginary axis to increase the range of accuracy of the AC procedure without losing the global $\mathcal{O}(N^4)$ scaling of the whole calculation. 

The merging of the contour deformation and analytic continuation of $W$ following Eq.~\ref{anacon0} is explained in the two following sections. We first introduce the specific quadrature along the imaginary axis for integration of the $GW$ self energy. The advantage of performing the AC on the screened Coulomb potential  to obtain efficiently and accurately the needed residues is then discussed.

\subsection{Quadrature for the imaginary-axis integral} 
\label{quadrature}

We start the presentation of our numerical treatment of Eq.~\ref{anacon0} by describing the quadrature we adopt for the calculation of the imaginary-axis integral contribution, namely the integral part $I(E)$ of $\langle \phi_m | \Sigma_C(E) | \phi_m \rangle $:
\begin{equation}
I(E) =\int   d\omega \; 
\langle \phi_m | G({\bf r},{\bf r}' ; E+ i\omega) \widetilde{W}({\bf r},{\bf r}' ;  i\omega)| \phi_m \rangle
\label{ImagContribution}
\end{equation} where the integral ranges from ${-\infty}$ to ${+\infty}$. One can rewrite  Eq.~\ref{ImagContribution} using once more the formal functional forms of $G$ and $\widetilde{W}$, leading to:
\begin{equation}
I(E)=\sum_{\lambda}\sum_{n}  I_{\lambda}^{ n}(E) \,| w_{\lambda}^{m,n} |^2  
\end{equation} with $w_{\lambda}^{m,n} = \langle \phi_m \phi_n  | w_{\lambda}  \rangle$ and 
\begin{equation}
\begin{split}
I_{\lambda}^{ n}(E)  = \int \! d\omega \, \frac{ 1 }{ {E+i\omega} - \varepsilon_n }  \left[  \frac{ 1}{ i\omega - \Omega_{\lambda} } 
 - \frac{ 1}{ i\omega + \Omega_{\lambda}  }  \right]
\end{split}
\end{equation} The value of $I_{\lambda}^{n}(E)$ depends on the sign of $(E-\varepsilon_n)$: 
\begin{equation}
\begin{split}
I_{\lambda}^{ n}(E)  = \left\{
\begin{array}{lr}
-\frac{2i\pi}{E-\Omega_\lambda-\varepsilon_{n} } & E<\varepsilon_{n}\\
\\
-\frac{2i\pi}{E+\Omega_\lambda-\varepsilon_{n} } & E>\varepsilon_{n}
\end{array}\right.
\end{split}
\end{equation}
Since the input energies $\varepsilon_n$ are known, contrary to the  $\Omega_\lambda$ poles of $\widetilde{W}({\bf r},{\bf r}' ; \omega)$, we can define an energy-specific quadrature $Q(\Delta E)=\big\{\omega_k,z_k(\Delta E)\big\}$ defined so as to minimise the error:
\begin{equation}
\int_{\Omega_{min}}^{\Omega_{max}} \!\!\!\!\!d\Omega\; \left|\left| 
\sum_k \left[  \frac{ z_k(\Delta E)}{ i\omega_k - \Omega } 
 - \frac{ z_k(\Delta E)}{ i\omega_k + \Omega  }  \right]  -  \frac{2i\pi}{\Delta E+\Omega }
\right|\right|
\end{equation}Using such quadrature, the imaginary axis integral contribution $I(E)$ can be estimated as 
\begin{equation}
I(E)=\sum_n \text{sgn}(\varepsilon_n - E ) I_n(E)
\end{equation}with
\begin{equation}
I_n(E)=\sum_k z_k(|E-\varepsilon_n|) \langle \phi_m\phi_n| \widetilde{W}(i\omega_k) | \phi_m\phi_n \rangle \
\end{equation}Such a contribution only requires the knowledge of $\widetilde{W}( i \omega )$ along the imaginary axis where it is smooth. The quadrature bounds $\Omega_{min}$ and $\Omega_{max}$ are adjusted depending on the system, typically setting  $\Omega_{min}$ as a fraction (a half in this work) of the input (e.g. Kohn-Sham) gap and  $\Omega_{max}$ as the maximum input ($\varepsilon_a - \varepsilon_i$) transition energy. 
Let's emphasise here that the $z_k(\Delta E)$ weight are computed ``on the fly" for each $\varepsilon _n$ contribution, while the $w_k$ sampling points are common to all these contributions and defined only once from the  $\Omega_{min}$ and $\Omega_{max}$ parameters. 

Anticipating on the results presented below, we collect in   Table~\ref{tab::convergence_MgO} the HOMO and LUMO energies obtained at the def2-QZVP $G_0W_0$@PBE level for the problematic MgO molecular system as a function of the number of $\lbrace \omega_k \rbrace$ quadrature frequencies along the imaginary axis, demonstrating a high level of accuracy even with a limited set of quadrature points. \\

\begin{table}[]
    \centering
    \begin{tabular}{rcccc} \\
 \hline
 n$\omega$ &  $E_{H}^1$ & $E_{H}^2$ & $E_{H}^3$ & $E_{L}$ \\
 \hline
 6 &  -6.6538(0.29)& -7.0835(0.19)&  -10.9733(0.10)  & -1.8752(0.80) \\
 8 & -6.6696(0.28) & -7.0943(0.20)&  -10.9787(0.10)  & -1.8907(0.80) \\
10 & -6.6711(0.28)& -7.0953(0.20)&  -10.9793(0.10)  & -1.8922(0.80)  \\
12 & -6.6712(0.28)& -7.0954(0.20)&  -10.9793(0.10)  & -1.8923(0.80)  \\
14 & -6.6714(0.28)& -7.0956(0.20)&  -10.9794(0.10)  &  -1.8924(0.80)  \\
 \hline
    \end{tabular}
    \caption{Convergence of the solutions of the quasiparticle equation for the HOMO and LUMO levels in MgO (in eV, with Z factors in parenthesis) as a function of the number n$\omega$ of $\lbrace \omega_k \rbrace$   quadrature frequencies along the imaginary axis. The three solutions with largest Z are provided in the case of the HOMO level. The uncertainties stemming from the residue calculations using AC techniques are cast away by using a large and constant number of reference frequencies within the complex plane (see text Section~\ref{continuation}).}
    \label{tab::convergence_MgO}
\end{table}

\subsection{Analytical Continuation of $W$} 
\label{continuation}

Although there are numerous techniques to do so (Pade approximants, AAA~\cite{AAA}, RKFit~\cite{RKFIT}, etc.) we choose here to perform the analytical continuation of the needed $W(\omega)$ residues: 

\begin{equation}
\widetilde{W}_{nm}( \omega ) = \langle \phi_m \phi_n | \widetilde{W}( \varepsilon_n - \omega ) | \phi_n \phi_m \rangle
\label{eq:residue}
\end{equation}

\noindent using the continued fraction approach proposed in Ref.~\citenum{Vidberg77}. Namely, given a set of reference points and associated values $\{(\omega_k,f_k)\}$, the domain of the corresponding function is extended to the complex plane by means of a continued fraction:
\[
f(z) := \frac{a1}{1+}\,\frac{a2(z-\omega_1)}{1+}\,\frac{a3(z-\omega_2)}{1+}\dots  
\]
This functional form is constructed recursively so that at stage $p$, the coefficient $a_p$ is set for $f(\omega_p)$ to take exactly the value $f_p$. Because of the $(z-\omega_{k})$ factors, adding a reference point $p$ doesn't affect the values of $f(\omega_k)$ for $k<p$. Such a form benefits in particular from recursion relations that allow  for fast calculations of $\{a_i\}$ coefficients and fast evaluations at any complex $z$ (see Ref.~\citenum{Vidberg77} Appendix for details). Anticipating on the two AC schemes presented below, we emphasise that there is no restriction on the location of the $\omega_k$ reference points that can be located anywhere in the complex plane. 

We slightly  adapted to the present problem the algorithm of Ref.~\citenum{Vidberg77} in the following way: i) since $W$ is symmetric in frequency, namely    $W(\omega)=W(-\omega)$, the corresponding functional form should be:
\begin{equation}
s(z) := \frac{a1}{1+}\,\frac{a2(z^2-\omega_1^2)}{1+}\,\frac{a3(z^2-\omega_2^2)}{1+}\dots , 
\end{equation}
the recurrence relations of Ref.~\citenum{Vidberg77} being modified accordingly;
ii) the inclusion order of the reference points in the continued fraction is determined so as to minimise the mean square error between the remaining references and continued expression at each recursion step; iii) a reference point is included in the continued fraction only if the AC result and the reference value differ by a significant amount ($>10^{-8}\times f_{\omega=0}$ here). In practice these last two considerations greatly stabilise the AC procedure and allowed us to work with large set of reference points without experimenting trouble with numerical stability.

\begin{figure}[ht]
\includegraphics[width=8cm]{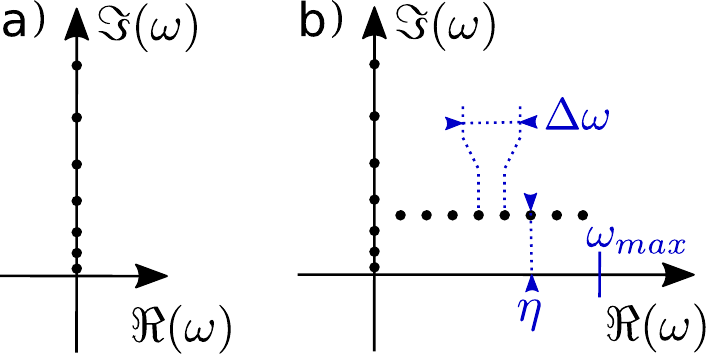}
\caption{Schematic representation of two possible frequency samplings (dots) over which the screened Coulomb potential $W$ is calculated explicitly  for the construction of an Analytic Continuation of $W(\omega)$ in the complex plane: a) along the imaginary axis only and b) with additional points parallel to the real axis. }
\label{fig_W_sampling}
\end{figure}

In this work, we thus propose to rely on the AC for the estimation of the   $\widetilde{W}_{nm}( E- \varepsilon_n)$ residues 
arising during the calculation of the $\langle \phi_m |\Sigma_C(E) |\phi_m \rangle $ correlation energy. For this, we experimented two different sampling strategies as exemplified in Figs.~\ref{fig_W_sampling}a-b, allowing to control the performance/accuracy ratio of the calculation.

The strategy of Fig.~\ref{fig_W_sampling}a considers reference points distributed along the imaginary axis only, taken identical to the points that were used to perform the numerical integration in Eq.~\ref{anacon0}. Once the screened Coulomb potential $W$ is constructed for a selected $\lbrace i \omega_k \rbrace$  grid along the imaginary axis, the construction of the $\widetilde{W}_{nm}$ values and their analytic continuations over the complex plane are obtained in negligible time. This matches the ``traditional" analytic continuation approach relying on data points calculated only at imaginary frequencies. As shown below,   very few sampling frequencies are necessary for an accurate analytic continuation of $W$ in the vicinity of the imaginary axis, allowing to get accurate quasiparticle energies for states relatively close to the energy gap. Following the convergence data reported in Table~\ref{tab::convergence_MgO} for the integral contribution to the self-energy, we will adopt in the following a limited n$\omega$=14 number of points along the imaginary axis. \\

On top of the imaginary axis grid points necessary to the numerical integration, the strategy of Fig.~\ref{fig_W_sampling}b also considers reference points distributed above the real axis in the vicinity of the  poles of $G(E+\omega)$ at energies ($E-\varepsilon_n$), where $E$ spans the range of energy for which $\Sigma(E)$ is needed.  As shown below, calculating explicitly $W(\omega)$ over a very coarse $\lbrace \omega_k \rbrace$ grid with typical spacing $\Delta \omega \simeq$ 1 eV  up to  $\Re(\omega_k) \le \omega_{max}$ allows to obtain very accurate AC expressions for $W(\omega)$ along (or close) to the real axis up to $\omega=\omega_{max}$, resulting in accurate self-energies $\Sigma(E)$ for $E$ within $\omega_{max}$ of the frontier orbitals. For sake of illustration, if one desires the quasiparticle correction to all valence states, with a valence bandwidth of about 30 eVs,   ($ \omega_{max} / \Delta \omega $) amounts to  30 additional $W$ matrices to be calculated. Due to the parity properties of $W$, the same calculated data points can be used to obtain the quasiparticle correction to all empty states within 30 eV from the gap. The same scheme will be adapted to core states in Section~\ref{sec:cores}.

The $\Delta \omega$ spacing can be also reduced to much smaller values  to check the convergence and acquire reference quasiparticle energies. We verified that a spacing of $\Delta \omega$=125 meV leads to a converged analytic continuation scheme.  This will be our strategy in the following to validate our calculations. In the following of the paper, such calculations will be referred to as our "reference" calculations with in parenthesis (Fig.~\ref{fig_W_sampling}b scheme, $\Delta \omega$=125 meV) as a reminder.

A parameter to be considered is the height $\eta$ of the additional grid points with respect to the real axis. Experimentation leads us to consider that $\eta=1.5 \times \Delta \omega$, with   $\Delta \omega$ the  grid spacing, was a reasonable value. We observed in particular that grid points lying too close to the real axis are detrimental to the AC precision, being probably  too sensitive to the local pole structure of $W(\omega)$. On the other hand, having the reference point too distant from the AC region would decrease the AC accuracy.

To conclude this section on the analytic continuation of $W$, we note that the scheme of Fig.~\ref{fig_W_sampling}b does not allow to get an accurate AC expression for $W(\omega)$ everywhere in the complex plane, but only in the ``vicinity" of the reference sampling points over which $W(\omega)$ is explicitly calculated.  One therefore cannot assume that the resulting AC of $W(\omega)$ will be accurate at much higher energy along the real-axis. As such, we did not attempt to use this AC form to perform explicitly the integral of the $G(E+\omega)W(\omega)$ integrand up to $\omega=+\infty$ along the real-axis following Eq.~\ref{sigma0}. In the contour deformation approach, $W$ is only required over a limited energy range, namely close to the ($\varepsilon_{n} - E$) poles of the $G(E+\omega)$ Green's function that entered the first/third quadrants.

\section{Technical details}

Benchmark calculations are first performed on the $GW$100 molecular test set.~\cite{Set15,Kra15,Car16,Mag17,Gov18,Wil18}
Molecular structures are taken directly from the $GW$100 original paper. \cite{Set15} 
Out of the full $GW$100 test set, we consider only the 93 systems not requiring pseudopotentials. 
For vynil-bromide and phenol, the corrected structures,~\cite{Mag17} and corresponding modified data,    
are taken from the $GW$100 website.~\cite{website}
Such a test set encompasses rare gaz  atoms, small diatomic molecules,   etc. up to  larger systems such as  the DNA nucleobases.
Our calculations are performed at the  def2-QZVP level \cite{Wei05} starting with input PBE \cite{Per96}  Kohn-Sham eigenstates to allow direct comparison 
with the available TURBOMOLE \cite{Set13} (TM) and FHI-aims \cite{Ren12} localized basis set calculations from Ref.~\citenum{Set15}.
The input Kohn-Sham calculations are obtained with the NWChem package.  \cite{NWCHEM} Coulomb-fitting resolution-of-identity (RI) \cite{Whi73} techniques are employed. 

We adopt the auxiliary def2-QZVP-RIFIT basis sets \cite{EMSL} that we consider to be better converged than the compact and  universal  Weigend Coulomb fitting
ones~\cite{Wei06} used in conjunction with the def2-QZVP Kohn-Sham basis in the TM-RI calculations of Ref.~\citenum{Set15}. 
Following a recent exploration of resolution-of-identity techniques, auxiliary basis sets are considered in their Cartesian representation, leading to
more accurate values as compared to the same basis set used in a spherical representation.~\cite{Duc17}
Core levels are included in the calculation of the susceptibility. 

The second molecular test set we explore is composed of 24 organic acceptors as proposed by Marom and coworkers.~\cite{Kni16} 
Such molecules have significant importance for e.g. hole-doping purposes and their large electronic affinity (EA) allows reliable measurements.
For comparison with available CCSD(T) data, \cite{She16} we adopt here the augmented \textit{aug}-cc-pVTZ basis sets together with their corresponding auxiliary basis sets \textit{aug}-cc-pVTZ-RI.\cite{EMSL} We compare in particular the ionization potential (IP) and electronic affinities (AE) obtained at the non-self-consistent $G_0W_0$@PBE0 level and ``gap-self-consistent" eg-$GW$@PBE0 approach with the available CCSD(T) data. 
Calculations are performed with the newly developed BeDeft program (BeyondDFT) that originates from rewriting and extending   the $GW$ and Bethe-Salpeter {\sc{Fiesta}} code. \cite{Jac15,Li16,Duc18}

\section{Results}

\subsection{ The $GW$100 test set }
\label{subsec:GW100}

We now explore the merits of performing the analytic continuation on the screened Coulomb potentiel $W$ for the residues in the contour-deformation scheme by studying the def2-QZVP $G_0W_0$@PBE quasiparticule HOMO and LUMO energies of the $GW100$ test set. We thus calculate  the quasiparticle energies obtained by the AC of $W$ starting from a very limited (n$\omega$=14) set of $W(\omega)$ matrices calculated at imaginary frequencies (scheme Fig.~\ref{fig_W_sampling}a) that we compare with the accurate reference calculations (Fig.~\ref{fig_W_sampling}b scheme, $\Delta \omega$=125 meV). The data are compiled in the Supporting Information  (Tables S1 and S2). 

The largest deviation with respect to the reference calculations  amounts to 6 meV for the  He atom HOMO for which the quasiparticle correction to the input Kohn-Sham energy is the largest ($\simeq$ 8 eVs). For such states with very large $GW$ correction, the $(E-\varepsilon_n^{KS})$ energies at which residues must be calculated, with $E\simeq E_m^{GW}$ the targeted quasiparticle energy, are lying far away from the imaginary axis. Besides He, 6 molecules show a discrepancy of the order of 1 to 2 meVs, all other systems displaying a deviation below the meV.  In the case of the AEs, one system (He again) presents a 1~meV deviation.   For sake of completeness, we also provide in the SI the def2-QZVP $G_0W_0$@PBE0 data that display even smaller differences between the two schemes due to  reduced $GW$ corrections. With a PBE0 starting point, the He HOMO displays the only discrepancy larger than a meV in relation again with a  $GW$ correction that is still substantial ($\simeq$6 eV). 

To further analyse the remarkable success of the present analytic continuation approach based on very few calculated data points along the imaginary axis, we now focus on the BN, O$_3$, BeO and MgO systems that were leading to difficulties when using the analytic continuation directly on the final self-energy matrix elements. Table~\ref{table2} reports the corresponding ionization potential (IP) taken as the negative of the HOMO def2-QZVP $G_0W_0$@PBE value. Our data are compared to the published values obtained with the Turbomole  and  FHI-Aims codes. \cite{Set15} We report the TM calculations performed with full diagonalization of the (RPA) dielectric matrices and no resolution-of-the-identity approximation, namely exact calculations that we label TM-noRI. The AIMS-16 and AIMS-P128 data were based on the analytic continuation of the self-energy using N=16 and N=128 parameters in the Pad\'{e} expansion of the self-energy matrix elements. 
Our results are given in the three last lines, using the notation beDeft-ref for the reference calculations (Fig.~\ref{fig_W_sampling}b scheme, $\Delta \omega$=125 meV) and beDeft-P14 for the efficient (Fig.~\ref{fig_W_sampling}a scheme, n$\omega$=14)  approach. The renormalization Z factor is defined as $\; 1/Z = 1 - \partial \Sigma(\omega) / \partial \omega \;$ where the derivative is taken at the quasiparticle energy.

\bgroup
\begin{table*}[t]
    \centering
    \begin{tabular}{l@{\hspace{2em}}|@{\hspace{2em}}c@{\hspace{2em}}c@{\hspace{2em}}c@{\hspace{2em}}c} & BN & $O_3$ & BeO & MgO \\
 \hline
AIMS-P16   & \phantom{11.}- -/\textbf{11.15} & 11.96/11.39            &  \textbf{9.07}/- -\phantom{7.} & \textbf{6.79}/- -\phantom{7.}\\ 
AIMS-P128  & 11.67/\textbf{11.03}            & 11.96/11.39            & \textbf{9.63}/8.58             & \textbf{7.11}/6.68           \\
TM-noRI    & $\phantom{^*}$11.67/\textbf{11.01$^*$}            & \textbf{11.95}/11.39   & \textbf{9.63}/8.62             & 7.09/\textbf{6.66}           \\
beDeft-ref & 11.667/\textbf{11.010}          & \textbf{11.967}/11.391 & \textbf{9.633}/8.608           & 7.096/\textbf{6.671}         \\
beDeft-P14 & 11.671/\textbf{11.010}          & \textbf{11.967}/11.392 & \textbf{9.635}/8.608           & 7.096/\textbf{6.671}         \\
Z          & 0.28/0.48                       & 0.40/0.25              & 0.47/0.15                      & 0.20/0.28                    \\
    \end{tabular}
    \caption{Ionization potential for BN, $O_3$, BeO and MgO taken as the negative of the HOMO quasiparticle energy at the def2-QZVP $G_0W_0$@PBE level. The AIMPS-P16, AIMS-P128 and TM (no-RI) data are taken from Ref.~\citenum{Set15}. The TM data should be considered as the reference def2-QZVP $G_0W_0$@PBE values. For each system and method, we provide when available the "left/right" solutions as defined in the Table 4 of Ref.~\citenum{Set15} and in the main text. The bold values select the true quasiparticle energies associated with the largest Z value. The three last lines correspond to the present study, with the label ``P14" for Fig.~\ref{fig_W_sampling}a scheme and ``ref" for Fig.~\ref{fig_W_sampling}b scheme with $\Delta \omega$=125 meV. Z values at second decimal place do not change between ref. and P-14 calculations. Energies are in eV. 
    $^*$We take the value from the $GW$100 website that differs by 10 meV from the Table 4 value of Ref.~\citenum{Set15}.}
    \label{table2}
\end{table*}
\egroup

\begin{figure}[h]
\includegraphics[width=7cm]{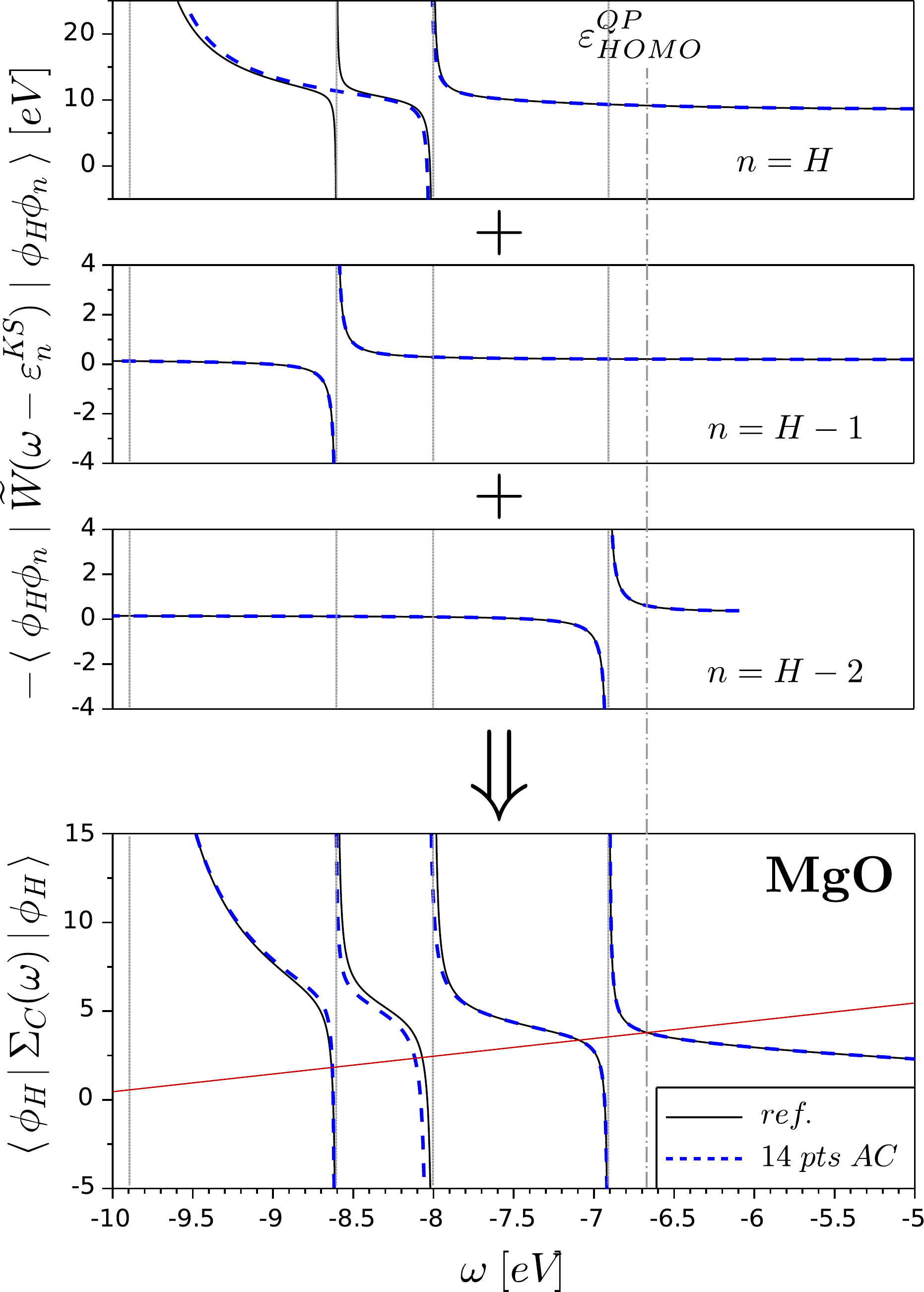}
\caption{ (Upper panels) Energy dependence  of the $\widetilde{W}_{nH}=\langle \phi_H \phi_n | W({\bf r},{\bf r}';\omega-\varepsilon_n) |  \phi_n \phi_H \rangle$ residues for MgO. The  wavefunction  $\phi_H$ is the PBE  HOMO eigenstate while the $\lbrace \varepsilon_n , \phi_n \rbrace$  with $n=(H-1,H-2)$ are the corresponding  HOMO-1 and HOMO-2  eigenstates. These contributions are compared to the overall $\langle \phi_H | \Sigma_{C}(\omega) | \phi_H \rangle$ correlation matrix element (Lower panel). The AC data obtained with n$\omega$=14 imaginary frequency data points  (dashed blue lines) are compared to reference calculations (full black lines) obtained with the fine ($\Delta \omega$=125 meV) sampling of  Fig. 2b.  The red line  represents the $f(\omega) = \omega - \varepsilon_H  - \langle \phi_H | \Sigma^X - V^{XC} | \phi_H \rangle $ function. The position of the $G_0W_0@PBE$ HOMO quasiparticle energy is indicated by the vertical black dashed line. }
\label{fig3}
\end{figure}

\begin{figure}[h]
\includegraphics[width=7cm]{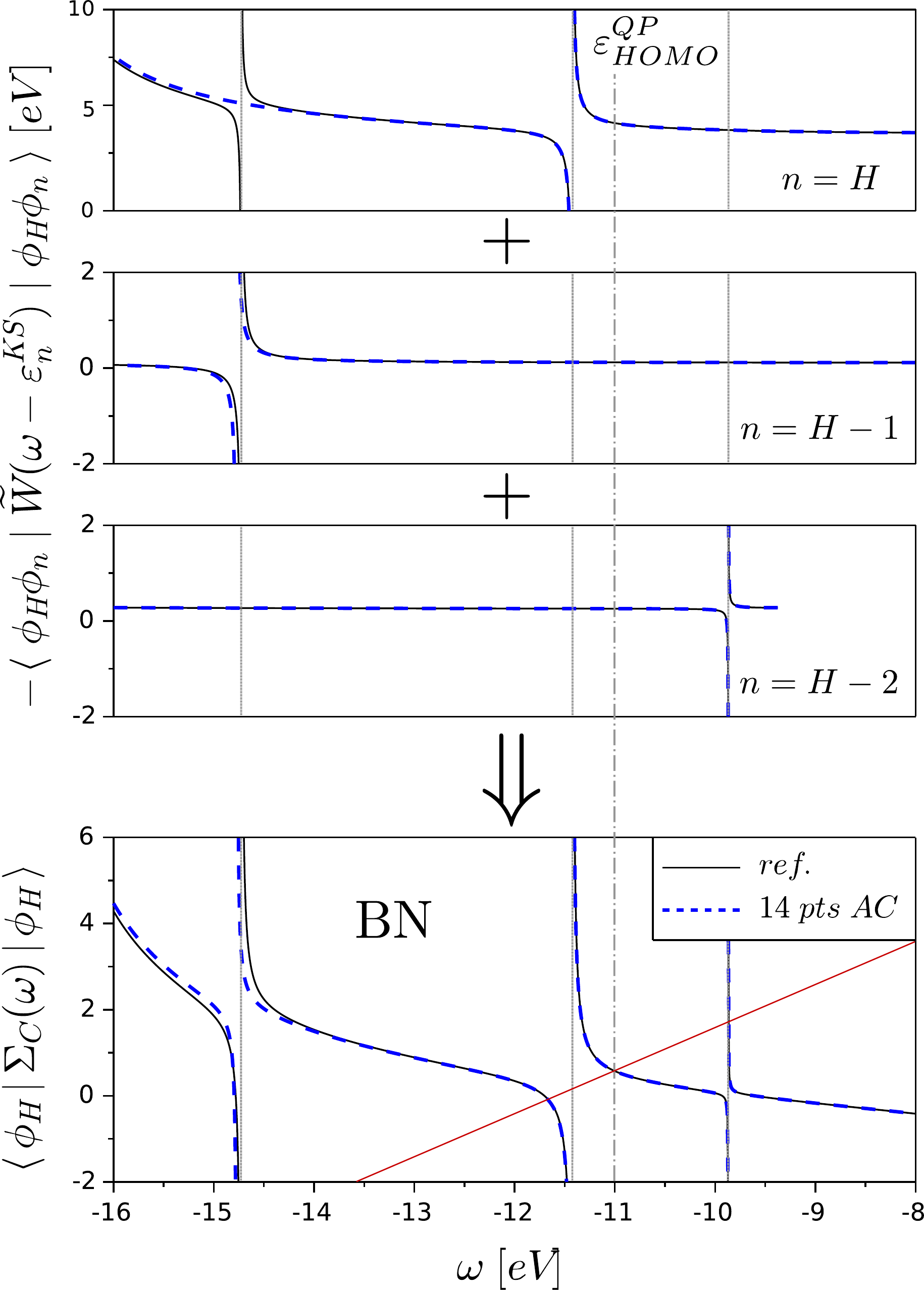}
\caption{ As in Fig.~\ref{fig3} but for the BN dimer. } 
\label{fig4}
\end{figure}

For each system and approach, we provide when available two values associated with the largest Z factors. The existence of several possible solutions of the quasiparticle equation:
\begin{equation}
 \varepsilon_n^{QP} = \varepsilon_n^{KS} + \langle \phi_n | \Sigma^{GW}(\varepsilon_n^{QP}) - V^{XC} | \phi_n \rangle
\end{equation}
is illustrated in  Figures~\ref{fig3} and \ref{fig4} (lower panels) for the specific MgO and BN cases. The straight red lines represent the $f(\omega) = \omega - \varepsilon_H^{KS} - \langle \phi_H | \Sigma^X - V^{XC} | \phi_H \rangle $ functions where (${\varepsilon}_H^{KS}, \phi_H$) are the PBE Kohn-Sham HOMO eigenstates. The possible quasiparticle energies are found at the crossing of such lines with the  $\langle   \phi_H | \Sigma^C(\omega) | \phi_H \rangle$ energy-dependent correlation self-energy expectation value (full black and dashed blue lines). The selected values with the largest Z factors correspond to the two solutions yielding the smallest IP.  Such solutions are labeled the -left- and -right- solutions in Ref.~\citenum{Set15}. \cite{centersolution} 

The analysis of the data in Table~\ref{table2} clearly indicates that our results (beDeft-P14) are within 20 meV of the reference TM-noRI HOMO values. For the present systems, the error associated with the (P14) analytic continuation is smaller than 2 meV (BeO) for the HOMO quasiparticle energies. As a result most of the discrepancy with TM-noRI data originates from the resolution-of-the-identity  approximation. In comparison, the AIMS-P16 approach could not identify all possible solutions, yielding further errors of the order of 120 meV, 560 meV and 320 meV for BN, BeO and MgO, respectively. Increasing the number of Pad\'{e} parameters up to 128 (AIMS-P128), requiring thus the calculation of the self-energy matrix elements at 128 imaginary frequencies, could restore  excellent results. However, the AIMS-P128 calculations does not identify the correct solutions for the O$_3$ and MgO systems, namely does not lead presumably to an accurate enough determination of the Z-factors. 
Such an analysis emphasizes the difficulties associated with techniques attempting to perform an analytic continuation of the self-energy matrix elements, as compared  to the present scheme where we perform an analytic continuation on individual screened Coulomb potential matrix elements.  

To better understand the improved stability of the present scheme, we now analyse in Fig…~\ref{fig3}-(Upper panels) the individual $\widetilde{W}_{nH}(\omega)$ matrix elements (see Eq.\ref{eq:residue}) with $n=(H,H-1,H-2)$ indexing the  MgO HOMO,   HOMO-1, and HOMO-2  eigenstates. Such states are selected by the two Heaviside functions in the contour-deformation formula. The reference $\widetilde{W}_{nH}(\omega)$ matrix elements (Fig.~\ref{fig_W_sampling}b scheme, $\Delta \omega$=125 meV) are plotted with  black solid lines. As discussed above, each individual $\widetilde{W}_{nH}(\omega)$ matrix element presents much less structures than the corresponding $\langle \phi_H | \Sigma_C(\omega) | \phi_H \rangle$ correlation self-energy term represented in the lower panel. In particular, the pole close to the targeted MgO HOMO quasiparticle energy at -6.67 eV originates from the (n=H-2) $\widetilde{W}_{nH}(\omega)$ contribution  characterized by a single pole. As a result, the analytic continuation using only data points along the imaginary axis (Fig.~\ref{fig_W_sampling}a) does an excellent job with our limited (n$\omega$=14) imaginary frequencies grid, reproducing very faithfully the main pic close to $\varepsilon_H^{QP}$ quasiparticle energy and leading to an accurate determination of the quasiparticle energy. The same quality of fit is obtained for the (n=H-1) contribution capturing accurately the pole at -8.6 eV. As a matter of fact, the number of reference points along the imaginary axis over which $\chi_0(i\omega)$ is explicitly calculated can be reduced down to about n$\omega$=8 to retain the meV accuracy on the HOMO energy as shown previously in Table~\ref{tab::convergence_MgO}.

The ideal case of one-pole only per $\widetilde{W}_{nH}(\omega)$ matrix element cannot be a general situation as shown in the (n=H) case where 3 poles are observed in the reference calculation. In this specific case, the analytic continuation of $\widetilde{W}_{nH}(i\omega)$ only captures the two broader pics, missing the sharpest one that is associated with a very small $w_{\lambda}^{mn}$ weight. Clearly, the very same pole is captured in the case of (n=H-1) where it appears with a larger weight. As such, the resulting pole in the self-energy is well reproduced with only a small deficit of weight very close to the corresponding energy resonance. We note that the pole at about -3.4 eV reported in Fig.~13 of Ref.~\citenum{Set15} originates from the n=LUMO $\widetilde{W}_{nH}$ contribution. \\

We now turn in Fig.~\ref{fig4} to the case of the BN dimer for which a 128-Pad\'{e} fit was required when attempting to perform a direct analytic continuation of the self-energy. The situation is similar to that of the MgO system with several poles dominating the $\langle \phi_H | \Sigma_C(E)  |  \phi_H  \rangle$ correlation energy dependence. However, an additional complication arises since the quasiparticle energy is related to the second pole of $\langle \phi_H | \Sigma_C(E)  |  \phi_H  \rangle$ (going away from the gap). Again, the poles of the self-energy are distributed amongst various $\widetilde{W}_{nH}$ matrix elements. Remarkably, the analytic continuation with a limited n$\omega$=14 imaginary axis sampling reproduces very accurately the sharp structure at -9.9 eV. Such a very good fit is made possible by the fact that it is the only pole of the (n=H-2) contribution. One pole at -15.6 eV is missed in the (n=H) contribution but it is captured by the (n=H-1) contribution, appearing with good accuracy in the overall self-energy dispersion.

We summarize this Section by concluding that distributing the poles of the self-energy amongst several $\widetilde{W}_{nm}( \omega )$ functions that are individually continued significantly stabilizes the analytic continuation approach, resolving in particular all the difficulties encountered for the $GW100$ test set with  a remarkably compact n$\omega$=14 energy-grid along the imaginary axis. We note that this grid was optimized for the integral contribution along the imaginary axis, and not for the analytic continuation. Additional points along the imaginary axis may potentially lead to a better capture of poles in the contributing $\widetilde{W}_{nm}( \omega )$ residues but without improving our results for the $GW100$ test set.

\subsection{ Acceptor molecules with large EA }
\label{subsec:acceptors}

We now turn to the set of Ref.~\citenum{Kni16} that contains 24 moderate size organic molecules relevant to organic electronics and light emitting devices.  We compile in Table~\ref{tab:marom_set_details}   our    $G_0W_0$@PBE0 values for the IP and AE energies. Our calculations are performed at the   aug-cc-pVTZ level for which selected reference CCSD(T) data \cite{She16} are available.
As now well documented, non-self-consistent $G_0W_0$ calculations starting with DFT Kohn-Sham eigenstates generated with limited amount of exact exchange (25$\%$ in the PBE0 case) lead to a significant underestimation of IP and AE values. The mean signed error (MSE) for the IP and AE amount to -0.32 eV and 0.32 eV, respectively, with maximum errors of -0.41 eV (dichlone IP) and 0.44 eV (TCNE AE). 

As compared to our reference $G_0W_0$@PBE0 calculations (Fig.~\ref{fig_W_sampling}b scheme, $\Delta \omega$=125 meV), the results obtained with the AC from n$\omega$=14 data points along the imaginary-axis yield  a maximum error of the order of 10$^{-2}$ meV over the entire set, demonstrating again the robustness of the present AC  scheme. This error is even smaller than that obtained with the $GW100$ test set, presumably because the $GW$ corrections are much smaller in the present case. This results from the PBE0 starting point (instead of PBE) and the larger size of the systems considered. Such considerations indicate  that the energies $(E-\varepsilon_n^{KS})$ at which the residues must be calculated are smaller, namely closer to the imaginary axis. \\

\begin{table}[h!]
\begin{tabular}{lrrrrrr}
& \multicolumn{2}{c}{CCSD(T)} & \multicolumn{2}{c}{$G_0W_0$}  & \multicolumn{2}{c}{eg$GW$} \\ 
& IP  &  EA & IP & EA & IP & EA \\
\hline
Anthracene                         &     7.47    &    0.26    &     7.16    &     0.56    &      7.30    &     0.37   \\
Acridine                           &       N/A    &      N/A    &     7.64    &     0.90    &      7.81     &     0.72   \\
Phenazine                          &     8.42    &    1.03    &     8.05    &     1.32    &      8.25    &     1.15   \\
Azulene                            &     7.49    &    0.48    &     7.20    &     0.72    &      7.36    &     0.54   \\
Benzoquinone                       &    10.17    &    1.46    &     9.81    &     1.75    &     10.41    &     1.54   \\
Naphthalenedione                   &     9.79    &    1.39    &     9.41    &     1.68    &     10.01    &     1.45   \\
Dichlone                           &     9.89    &    1.82    & {\bf 9.48}  &     2.12    &      9.72     &     1.90   \\
F4-benzoquinone            &    11.04    &    2.18    &    10.68    &     2.48    &     10.95    &     2.29     \\
Cl4-benzoquinone            &    10.12    &    2.36    &     9.79    &     2.67    &     10.03    &     2.46   \\
Nitrobenzene                       &    10.14    &    0.44    &     9.85    &     0.75    &     10.06     &     0.48   \\
F4-benzenedicar.   &    10.66    &    1.51    &    10.32    &     1.84    &     10.57    &     1.67   \\
Dinitrobenzoni.                &    11.07    &    1.67    &    10.81    &     2.02    &     11.06     &     1.75   \\
Nitrobenzoni.                  &    10.55    &    1.20    &    10.26    &     1.56    &     10.49     &     1.32   \\
Benzonitrile                       &     9.88    &   -0.29    &     9.57    &    -0.02    &      9.78     &    -0.24   \\
Fumaronitrile                      &    11.40    &    0.89    &    11.03    &     1.21    &     11.28     &     0.98   \\
mDCNB                              &    10.37    &    0.54    &    10.08    &     0.86    &     10.30     &     0.63   \\
TCNE                               &    11.90    &    2.94    &    11.52    &     3.34    &    
11.77     &     3.17   \\
TCNQ                               &     9.49    &    3.23    &     9.20    & {\bf 3.67}  &      9.35    & {\bf 3.56} \\
Maleicanhydride                    &    11.27    &    0.92    &    10.85    &     1.15    &     11.45    &     0.92   \\
Phthalimide                        &    10.02    &    0.54    &     9.75    &     0.79    & {\bf 10.36}  &     0.56   \\
phthalicanhydride                  &    10.48    &    0.78    &    10.18    &     1.04     &     10.39    &     0.81   \\
Cl4-isobenz.    &       N/A    &      N/A    &     9.65    &     1.84    &      9.87    &     1.61    \\
NDCA                               &       N/A    &      N/A    &     8.78    &     1.46    &      8.94    &     1.27   \\
bodipy                             &       N/A    &      N/A    &     7.88    &     1.76    &      8.00    &     1.61   \\
\hline
MaxErr   &      &      &  -0.41  &  0.44  & 0.34 & 0.33 \\
MAE      &       &      &  0.33  & 0.31   & 0.14 & 0.10 \\
MSE      &       &      &  -0.33 & 0.31 & -0.04 & 0.10 \\
\hline
\end{tabular}
\caption{Ionization potential (IP) and electronic affinities (AE)   aug-cc-pVTZ $G_0W_0$@PBE0 data for  Ref.~\citenum{Kni16} test set. The CCSD(T) data are from Ref.~\citenum{She16}. The IP and AE yielding the largest discrepancy (Max.Err.) with the CCSD(T) data are highlighted. The Maximum error (MaxErr), the mean absolute error (MAE) and mean signed error (MSE) are given with respect to CCSD(T). The same data are reproduced in the SI at the third decimal place for reference. We further provide for information the data associated with the simple ``gap-self-consistent" (eg$GW$) calculations. }
\label{tab:marom_set_details}
\end{table}

While various schemes were explored to improve  $G_0W_0$@PBE0  data for this set of molecules, \cite{Kni16} including  optimal DFT starting points generated with  tuned range-separated hybrids, \cite{Gal16} self-consistency on both eigenvalues and eigenstates (sc$GW$), or by including more diagrams such as the second-order screened exchange (SOSEX) term, we explore here a very simple approach, namely the partially self-consistent eg$GW$ scheme where only the value of the HOMO-LUMO gap is updated (-eg- stands for energy gap) in the spirit of a self-consistent  scissor approximation. This is the simplest self-consistent scheme for approaches providing at reduced cost the energy of frontier orbitals.  The eg$GW$ calculations reduce  the mean absolute error (MAE) on the IP from 0.329 eV to 0.135 eV, as compared to the single-shot $G_0W_0$@PBE0 technique, and from 0.307 eV to 0.098 eV in the case of the AE.




\section{ Beyond frontier orbitals } 

The eg$GW$ scheme is a simplification of the popular ev$GW$ approach \cite{Bla11,Ran16,Kap16,Gui18} where all eigenvalues are self-consistently updated. While the eg$GW$ approach only requires calculating the HOMO and LUMO quasiparticle energies, the full ev$GW$ approach requires calculating the energy of states located far away from the gap. 
Beyond the ev$GW$ approach, accessing reliably specific levels located far away from frontier orbitals is an important issue. One could invoke e.g. core states \cite{Set18,Gol18} or the frontier orbitals of molecular systems deposited on metallic electrodes that are separated from the Fermi level by the ``continuum" of metallic states. \cite{Nea06,Thy09} This is a severe test for analytic continuation techniques relying on data acquired only along the imaginary axis as we now discuss. 

To illustrate the difficulty associated with deeper states,
we extract from the previous set of moderate size molecules the paradigmatic  TCNQ electron acceptor that belongs to an important family (F4TCNQ, F6TCNNQ, etc.)  of \textit{p}-type dopants in organic semiconductors. We select for illustration the n=HOMO-35th state located  $\simeq$18.5 eV below the gap ($G_0W_0$@PB0 value). We start again with reference calculations by using a very large set of energy grid points close to the real axis (Fig.~\ref{fig_W_sampling}b scheme, $\Delta \omega$=125 meV) on which $W$ is explicitly calculated for the construction of an accurate AC to the complex plane. 

\begin{figure}[h]
\includegraphics[width=8cm]{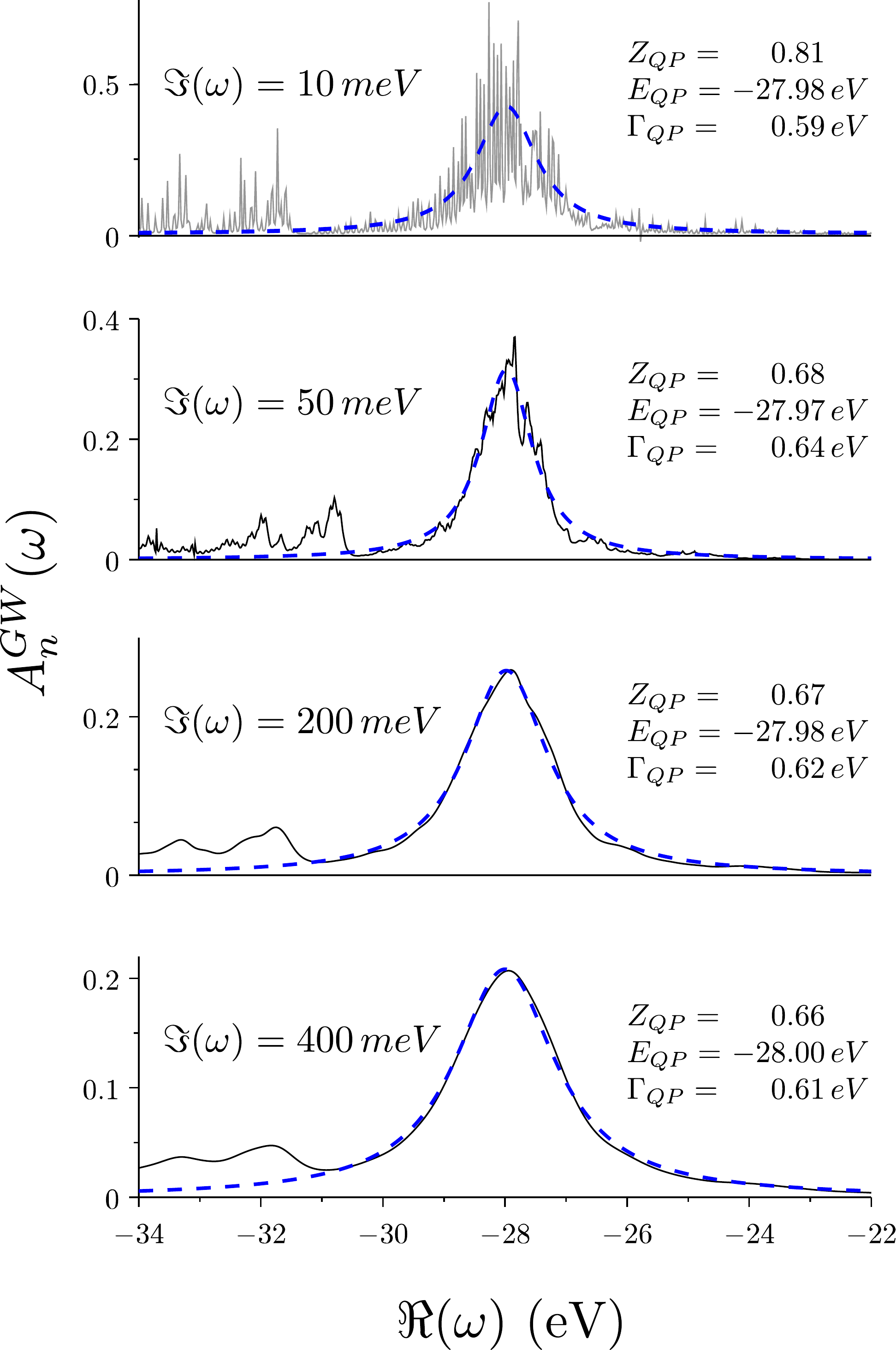}
\caption{ Spectral function $A_n^{GW}(\omega)$ associated with the TCNQ n=HOMO-35 state located at about 18.5 eV below the gap ($G_0W_0@PBE0$ value). The spectral function is calculated along paths parallel to the real axis with a constant $\Im(\omega)$ imaginary part of 10, 50, 200 and 400 meV. The blue dashed lines are the one-pole $A_n^{fit}(\omega)$ Lorentzian fit to the spectral function used to extract the quasiparticle energy $E_{QP}$, lifetime ${\Gamma}_{QP}$ and renormalization factor $Z_{QP}$. }
\label{TCNQ_HOMO-35}
\end{figure}

We plot in Fig.~\ref{TCNQ_HOMO-35} the corresponding spectral function :
$$
A_n^{GW}(\omega) = \frac{1}{\pi} \Big| \Im  \Big( \langle \phi_n |  G(\omega)  | \phi_n \rangle  \Big) \Big|
$$
with $\;  G = G_0 + G_0 (\Sigma^{GW} - V^{XC} ) G \;$ and $G_0$ the input PBE0   Green's function. More specifically, we represent $A_n^{GW}(\omega)$ along $\omega$-lines parallel to the real-axis  with a constant imaginary part $\Im({\omega})$ ranging from 10 meV to 400 meV. For very small  $\Im({\omega})$ value (10 meV), namely close to the real axis, the spectral function exhibits a very dense ``forest" of peaks corresponding to a large density of solutions of the quasiparticle equation, all solutions associated with a very small Z factor. The presence of this very large density of poles for the self-energy in this energy range originates from the large number of poles of the susceptibility associated with transitions from occupied states to the quasi-continuum of empty states above the vacuum level. 

Any analytic continuation from the imaginary axis will experience great difficulties to describe such a complex structure along (or very close) to the real axis. We emphasize that the poles of $W$ in this energy range becomes denser and denser as the size of the Kohn-Sham basis set increases, following the densification of empty states above the vacuum level. Eventually, all sharp contributions will merge into a broad structure forming the quasiparticle peak with a finite spectral weight.

This formation of a smooth quasiparticle peak can be achieved by increasing the $\Im({\omega})$ value, as shown in  Fig.~\ref{TCNQ_HOMO-35}. When $\Im({\omega})$ becomes sufficiently large, the quasiparticle peak clearly appears together with  the incoherent background at lower energy. 
To extract the quasiparticle energy, we fit the spectral function by a one-pole spectral (or Lorentzian) function:
\begin{equation}
\begin{split}
A_n^{fit}(\omega) =& \frac{ 1} {  \pi } \left|   \Im  \left(  \frac{ Z_{QP} }{  \omega - ( E_{QP} + i\Gamma_{QP} ) } \right) \right| \\
                  =& \frac{Z_{QP}}{\pi} \frac{ \big|\Im(\omega)- {\Gamma_{QP}} \big| }{ (\Re(\omega)- E_{QP} )^2 + (\Im(\omega)-{\Gamma}_{QP})^2 } 
                  \label{eqn:fit}
\end{split}
\end{equation}
resulting in the dashed-blue lines of Fig.~\ref{TCNQ_HOMO-35}. Together with the standard   quasiparticle energies $E_{QP}$ and renormalization factor $Z_{QP}$, one obtains the inverse lifetime $\Gamma_{QP}$. 
The lorentzian fit leads to a stable quasiparticle energy nearly independent of the $\Im(\omega)$ value. The  renormalization factor $Z_{QP}$ and inverse lifetime $\Gamma_{QP}$ are also found to be reasonably stable provided that $\Im(\omega)$ is chosen not too small. 

\begin{figure}
\includegraphics[width=8cm]{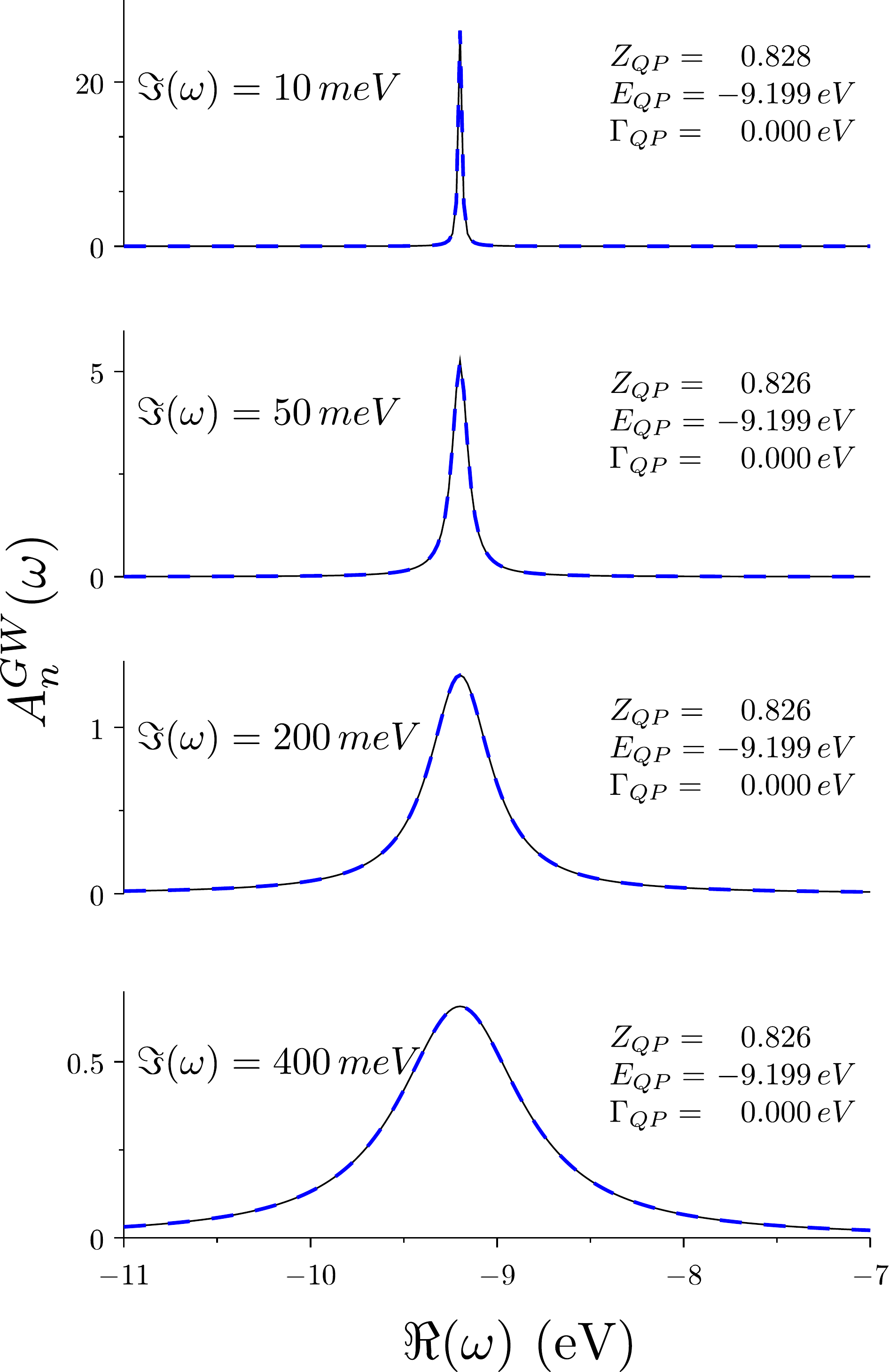}
\caption{ Same as Fig.~\ref{TCNQ_HOMO-35} but for the TCNQ HOMO spectral function. }
\label{SpFunc_HOMO-0}
\end{figure}

In the case of a well-defined quasiparticle peak (i.e. a single, well separated pole), such as the frontier orbitals discussed here above, this method yields the very same result as the one obtained by directly solving the quasiparticle equation (e.g. Fig.~\ref{fig3}). This is illustrated in Fig…~\ref{SpFunc_HOMO-0} in the case of the TCNQ HOMO level 
where the $E_{QP}$, $Z_{QP}$ and $\Gamma_{QP}$ are completely insensitive to the choice of $\Im(\omega)$. In such cases, the fit yields a vanishingly small  $\Gamma_{QP}$ inverse lifetime, as expected for states close to the gap.

The present analysis suggests that studying the spectral function $A_n^{GW}(\omega)$ along energy lines slightly shifted away from the real-energy axis, rather than attempting to capture all details of the self-energy along the real-axis, provides a much simpler and stable way to extract quasiparticle energies, together with meaningful renormalization factor $Z_{QP}$ and inverse lifetime $\Gamma_{QP}$. As exemplified here above, the obtained quasiparticle characteristics extracted through the one-pole fit strategy (Eq.~\ref{eqn:fit}) are very much insensitive to the choice of the $\Im(\omega)$ value. In the following, we adopt this strategy with $\Im(\omega)$ = 100 meV to evaluate the merits of the present analytic continuation approach. 

We thus perform the same analysis for all molecules of the set, considering all states within 20 eV from the gap (PBE0 value), encompassing namely over $600$ occupied and $2600$ unoccupied levels. As above, we first establish reference ($E_{QP}, Z_{QP}, {\Gamma}_{QP}$) values using  explicitly calculated $W$ matrices over a dense grid of energy points close to the real-axis (Fig.~\ref{fig_W_sampling}b scheme, $\Delta \omega$=125 meV). From this large data set, $W$ is analytically continuated to the $\Im(\omega)$= 100meV axis parallel to the real-axis along which the self-energy and spectral function $A_n^{QP}(\omega)$ are constructed. The Lorentzian fit of this accurate spectral function leads to reference quasiparticle data. 

\begin{figure}[ht]
 \includegraphics[width=7cm]{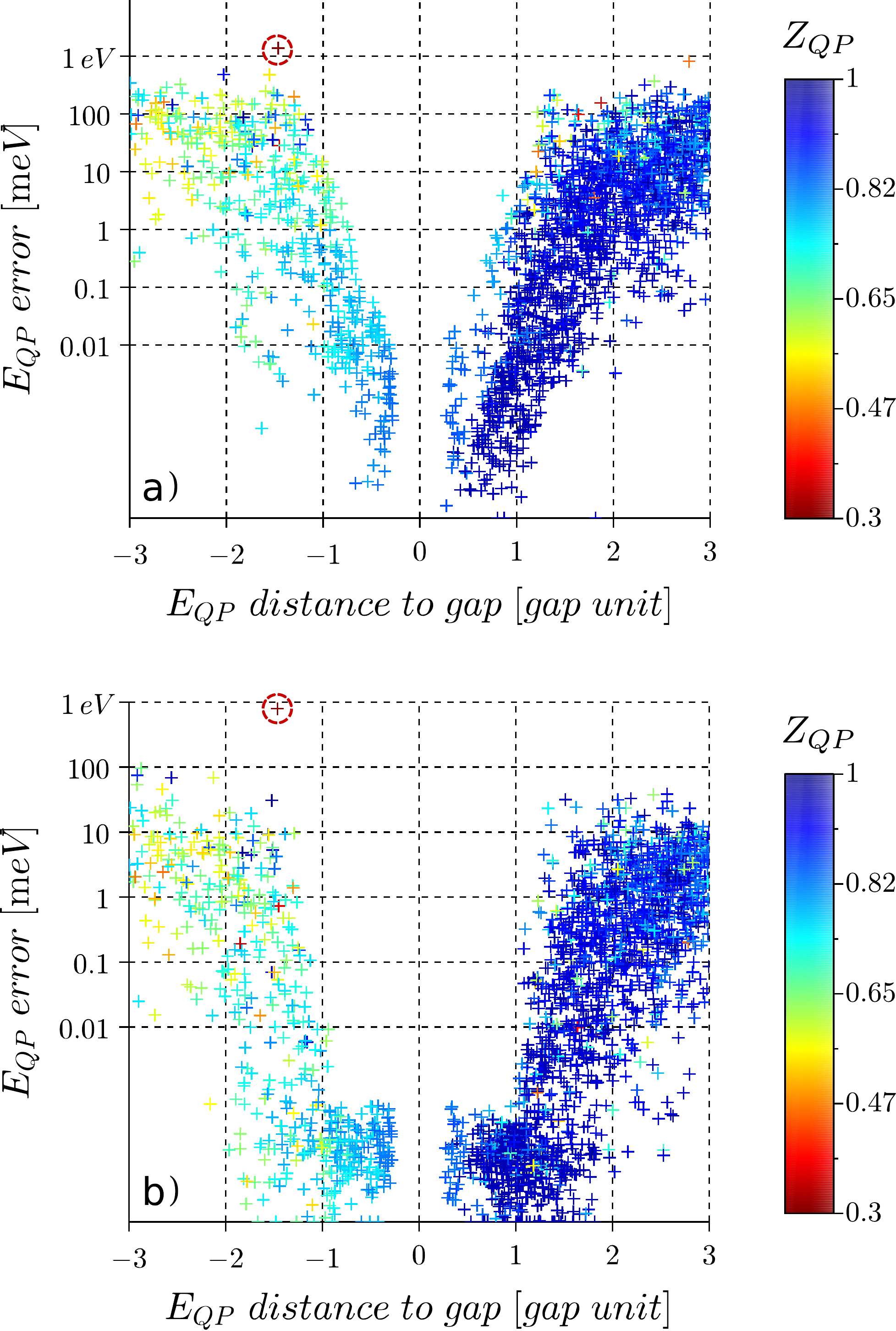}   
\caption{ Plot of the error induced by the analytic continuation as a function of the distance in energy of the considered states with respect to the gap edges (PBE0 input values) in units of the corresponding PBE0 gap. The color code indicates the Z-factor. Errors are given in meV in log scale. The graph compiles the data for all the molecules of the acceptors set (Ref.~\citenum{Kni16}). In (a), the analytic continuation of $W$ relies on the n$\omega$=14 data points calculated along the imaginary axis (see Fig.~\ref{fig_W_sampling}a). In (b) additional reference $W$ matrices are calculated over a very coarse energy-grid  parallel to the real-axis with a $\Delta \omega=$1~eV spacing (see Fig.~\ref{fig_W_sampling}b), resulting in a more accurate analytic continuation of $W$ to the $\Im(\omega)=100\,\mathrm{m}eV$ energy axis. The red cross with a circle identifies the main outlier over the entire set (see text). }     
\label{QP_error}
\end{figure}

We now perform analytic continuations of $W$ to the  $\Im(\omega)$=100meV axis starting from much reduced sets of explicitly calculated $W$ matrices.  We start by  the set of $W$ matrices calculated over the n$\omega$=14 integration points along the imaginary-axis as in Sections~\ref{subsec:GW100} and~\ref{subsec:acceptors}. After Lorentzian fit of the resulting spectral function $A_n^{QP}(\omega)$, the extracted quasiparticle energy $E_n^{QP}$ is compared to the reference values. We plot in Fig.~\ref{QP_error}(a)   the resulting error in meV (log scale).

For frontier orbitals, namely the HOMO and LUMO levels, the errors are vanishingly small ($\le$ 0.01 meV). This is consistent with the results of  Sections~\ref{subsec:GW100} and ~\ref{subsec:acceptors} revealing the accuracy of the analytic continuation of $W$ using few data points obtained along the imaginary axis only.  Further, this confirms that the extraction of the quasiparticle energies through the one-pole fit of the spectral function is stable and accurate for frontier orbitals.  Very consistently, recalculating the HOMO and LUMO energies  of the $GW$100 test set with the present scheme produces trifling errors as compared to the results of Section~\ref{subsec:GW100} where quasiparticle energies were obtained by solving the quasiparticle equation along the real axis. Similarly, we verify that the  quasiparticle spectral weight $Z_{QP}$ stemming from the one-pole fit is identical to the renormalization factor obtained previously from the energy derivative of the self-energy  at the quasiparticle energy.  

Apart from frontier orbitals, the errors remain  acceptable (below 10 meV) for states within one energy gap of frontier orbitals. In this energy range, the quasiparticle renormalization factor $Z_{QP}$ remains significant ($\geq 0.65$) indicating well defined quasiparticles. 
As such, the present analytic continuation scheme, relying on calculating  $W$ over a few points along the imaginary axis, is accurate not only for the HOMO and LUMO levels, but also for states located typically within one energy gap from frontier orbitals. As expected however, for states deeper in energy, the error increases, becoming occasionally close to an eV. 

We now increase the accuracy for deeper state by explicitly calculating $W$ over a very coarse grid of frequencies following the scheme of Fig.~\ref{fig_W_sampling}b but with a large $\Delta \omega=$1 eV energy spacing between grid points, leading to calculating $W$ matrices  over n$\omega \simeq 28$ additional points on top of the n$\omega$=14 points along the imaginary axis. The present scheme is thus intermediate between the "reference" calculations ($\Delta \omega=$125 meV)  and the analytic continuation of $W$ from the imaginary-axis-only data points. Contrary to the self-energy operator $\Sigma(\omega)$,   calculating the screened Coulomb potential $W(\omega)$ at frequencies with a large real-part  does not present additional difficulties as compared to calculating $W(\omega)$ on the imaginary axis, offering in particular the very same cost and $\mathcal{O}(N^4)$ scaling with system size.~\cite{analytic0}

The error associated with this intermediate scheme is reported in  Fig.~\ref{QP_error}(b). Clearly, adding $\simeq 28$ frequencies in the first-quadrant over which $W$ is explicitly calculated allows to significantly reduce the error within the $\pm20\,eV$s range explored. In particular, the error for states located within an energy gap of frontier orbitals decreases dramatically to negligible values. Further, the errors fall systematically below 0.1 eV within the $\pm3$ energy gap window.

\begin{figure}[ht]
 \includegraphics[width=8cm]{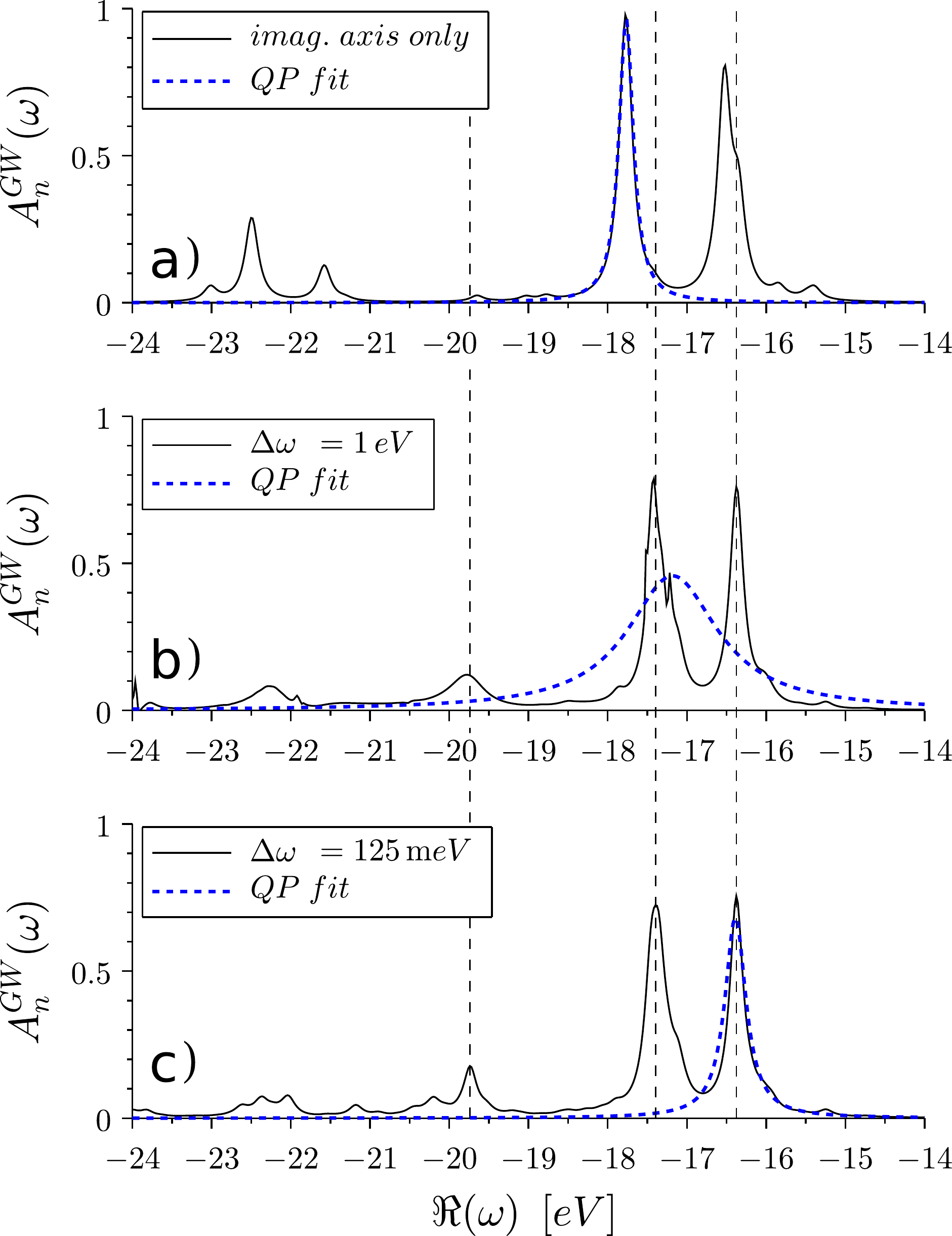}   
\caption{ Spectral function and one-pole quasiparticle (QP) fit associated the HOMO-15 of Nitrobenzonitrile, for different AC strategies: a) reference $W$ are calculated along the imaginary axis only; b) Fig.~\ref{fig_W_sampling}b scheme with $\Delta \omega=1\,eV$ grid spacing along real axis; c) Fig.~\ref{fig_W_sampling}b scheme with $\Delta \omega=125\,\mathrm{m}eV$ grid spacing along real axis (taken as reference calculation). The spectral function is plotted at $\Im(\omega)=100\,\mathrm{m}eV$.   
This illustrate a typical pathological case for the quasiparticle energy definition.}     
\label{nitrobenzo}
\end{figure}

\begin{figure}[h]
 \includegraphics[width=8cm]{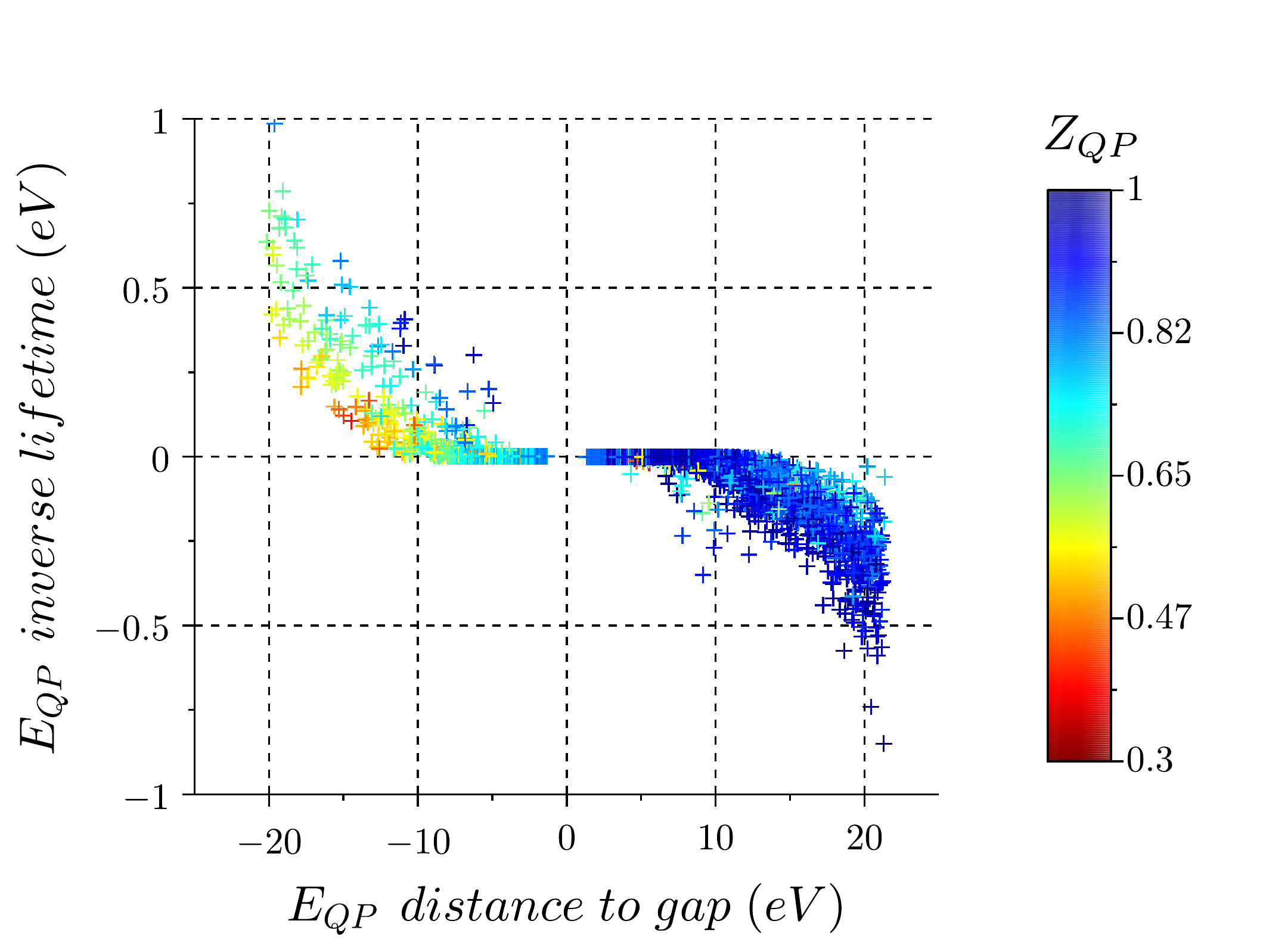}   
\caption{ Inverse quasiparticle lifetime $\Gamma_{QP}$ (in eV) as obtained from the one-pole quasiparticle fit of the spectral function.    }     
\label{QP_lifetime}
\end{figure}

Out of the $\simeq$600 occupied states considered, one clear outlier presents a large discrepancy between the quasiparticle energy associated with the $\Delta \omega=1\,eV$ coarse grid and the reference calculation (see red circled cross in Fig.~\ref{QP_error}). We plot in Fig.~\ref{nitrobenzo} the corresponding spectral function associated with the HOMO-15 state of nitrobenzonitrile located $\sim 6 eV$ below the gap ($G_0W_0$ value). We take this outlier as a good example of the limit of a one-pole fit procedure. As can be observed in Fig.~\ref{nitrobenzo} where the reference spectral function are provided, $A_n^{GW}(\omega)$ is dominated by two peaks of similar weight, each peak being characterized by a small Z-factor of about 0.3. 

In both the reference (Fig.~\ref{nitrobenzo}c) and imaginary axis only (Fig.~\ref{nitrobenzo}a) calculations, the one-pole fit selects the peak with the largest intensity. However, the AC from the imaginary axis   reproduces both peaks with significant errors regarding their energy location and intensity. As a result, the lower energy structure is erroneously chosen as the quasiparticle peak.  In the case of the analytic continuation based on calculating $W$ over the coarse $\Delta \omega$=1 eV  energy grid (Fig.~\ref{nitrobenzo}b), although the structures of $A_n^{GW}(\omega)$ are well-captured (see dashed vertical lines), the fit procedure fails to select a single peak and averages both structures with a very broad lorentzian. The resulting $Z$ factor exceeds unity, providing a good indication that the fit failed to identify the quasiparticle correctly. 

We conclude from the present example that for deeper states, the AC technique can be made accurate with additional $W$ reference points in the first-quadrant, even in situations where the spectral function is dominated by several structures. However, there may be difficulties in attempting to extract a well defined quasiparticle peak from the spectral function. Such an observation questions the concept of quasiparticle rather than the quality of the AC scheme. The error analysis for all occupied/unoccupied states is provided in the SI, showing no other outliers below the gap, and only 3 states out of $\simeq$2600 with an error larger than 0.1 eV in the unoccupied manifold at larger energy ($\simeq$ 4 gaps value). 

We finally plot in Fig.~\ref{QP_lifetime} the $\Gamma_{QP}$ (signed) inverse lifetime. As expected, we observe a significant increase as a function of the distance in energy from mid-gap, even though not following clearly the standard quadratic $(E-E_F)^2$ dependence associated with 3D Fermi liquids. \cite{Pines} Discussing whether the concept of quasiparticles remains valid for states located deep into the occupied/unoccupied manifolds is beyond the scope of the present paper concerned with the analytic continuation approach. We can only recommend for very low lying states to analyze the full spectral function that can be accurately captured by the analytic continuation of $W$ following the scheme of Fig.~\ref{fig_W_sampling}b, even in the limit of a very coarse $\Delta \omega$ sampling.  Too small $Z$ factors may be chosen as an indication that difficulties potentially exist.

\section{Core states}
\label{sec:cores}

\begin{figure}[ht]
 \includegraphics[width=8cm]{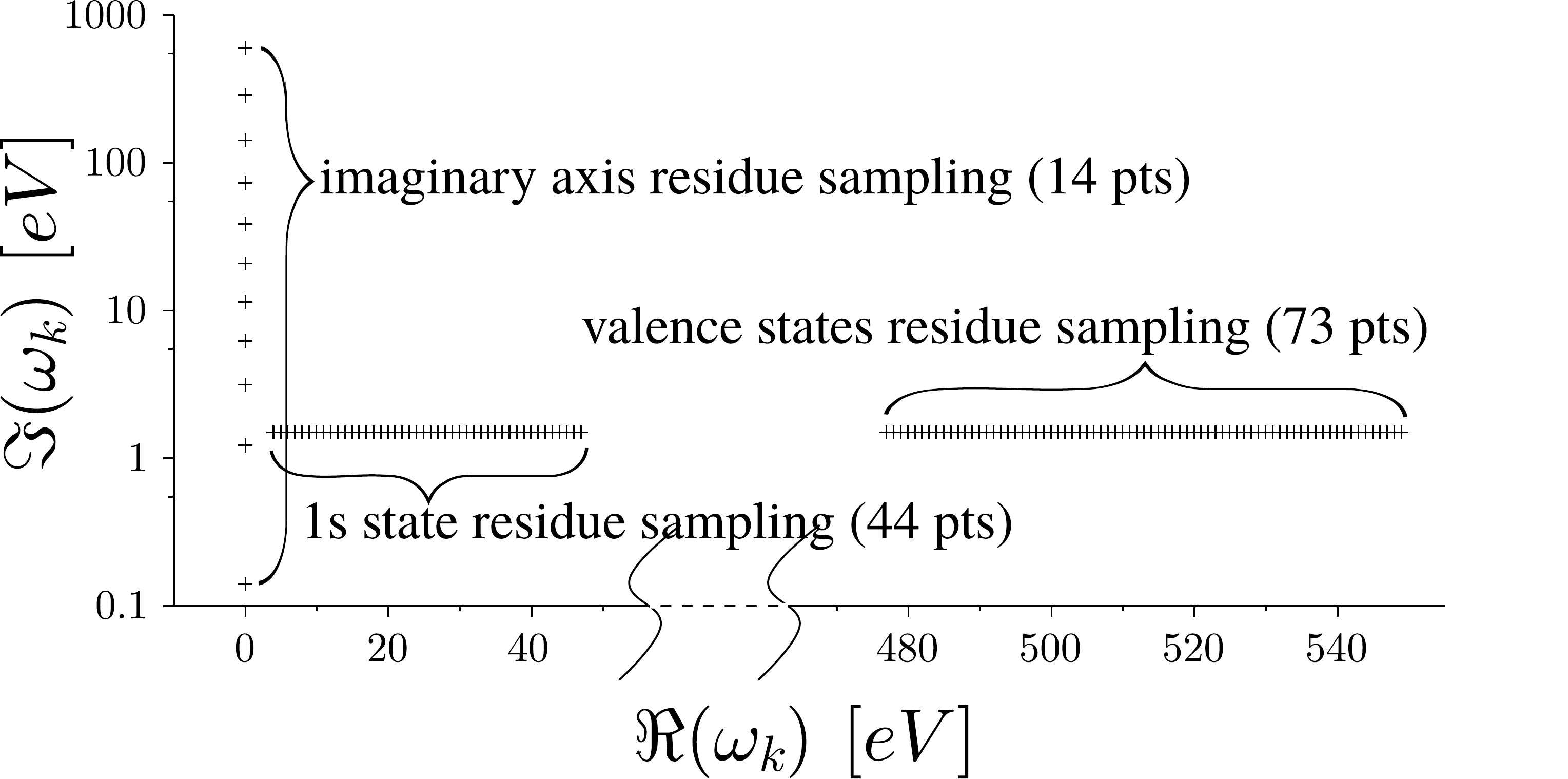}   
 \caption{1s core state $W$ sampling strategy for $H_2O$. The $\Delta \omega$ spacing above the real axis is $1\,eV$, leading to a total of 131 reference frequencies for the analytical continuation of $W$ residues required by the calculation of $\Sigma^C(\omega)$ within a large $50\,eV$ energy window.}     
 \label{water_core_sampling}
\end{figure}

We conclude this exploration by considering core states, focusing on the $H_2O$  1\textit{s} energy level as a test case. For such core levels, the residuals ${W}(\varepsilon_i - E_C)$, where $E_C$ is some typical core energy, can be partitioned into two categories, one with $\varepsilon_i$ a core level leading to small $|\varepsilon_i - E_C | $ energies and the other  with $\varepsilon_i$ a valence level leading to very large (several hundred eVs) $|\varepsilon_i - E_C | $ energies. Following our sampling strategy of Fig.~\ref{fig_W_sampling}(b), we verified that calculating $W(\omega)$  matrices for $Re(\omega)$ located in the large energy range between $|\varepsilon_i - E_C | $ for core and valence states does not influence the quality of the analytic continuation of $W$ where it is needed. This leads to the sampling scheme of Fig.~\ref{water_core_sampling} which is a variation on the sampling scheme of Fig.~\ref{fig_W_sampling}(b). In practice, with a $\Delta \omega$ spacing of 1 eV, we calculate explicitly 44 $W(\omega)$ matrices for $0 \le \Re(\omega) \le 45$ eVs to allow plotting   $\Sigma(E)$ on a large energy range around the 1\textit{s} core quasiparticle energy, while $\simeq$73 frequencies are kept for sampling the residues associated with valence states.

\begin{figure}[t]
 \includegraphics[width=6cm]{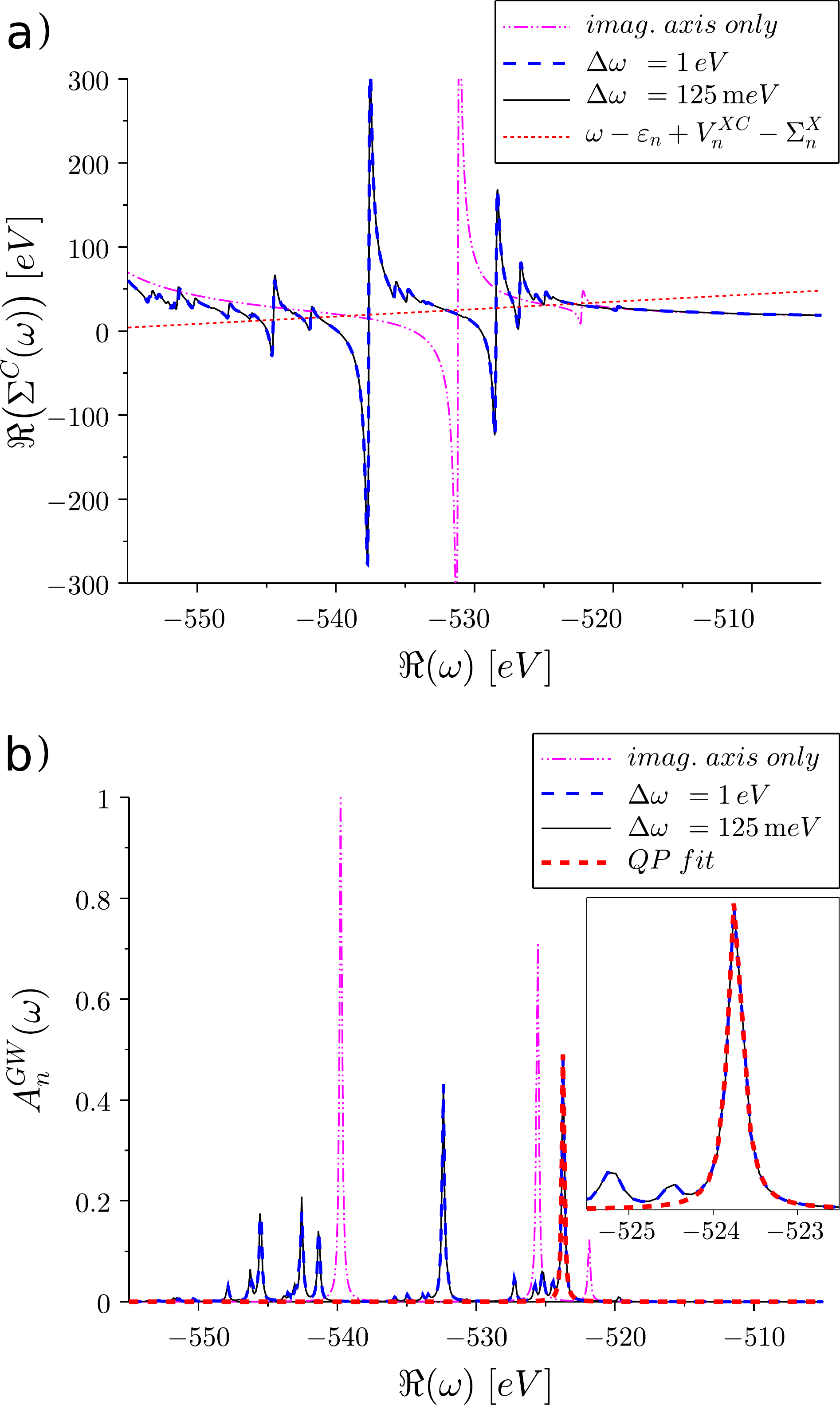}   
\caption{$GW$ calculation for the $\phi_{1s}$ H$_2$O core state: a) real part of the correlation self-energy $\langle \phi_{1s} | \Sigma^C(\omega) | \phi_{1s} \rangle$ expectation value; b) associated spectral function.  We compare: (full black lines) reference calculation obtained using Fig.~\ref{fig_W_sampling}b scheme with $\Delta \omega=125\,\mathrm{m}eV$; (dashed blue lines) calculation obtained using Fig.~\ref{water_core_sampling} scheme with $\Delta \omega=1\,eV$ grid spacing; (dashed pink lines) calculation obtained using Fig.~\ref{fig_W_sampling}a scheme with n$\omega$=14 points along the imaginary axis only. The correlation self-energies and spectral functions are plotted for $\Im(\omega)=100\,\mathrm{m}eV$. The Inset zooms the spectral functions around the quasiparticle energy.}     
\label{water_core}
\end{figure}

With such reference points, we analytically continue $W$ to the $\Im(\omega)$=100 meV line along which we plot the real-part of the correlation self-energy (Fig.~\ref{water_core}a) and the spectral function (Fig.~\ref{water_core}b). We chose the energy range considered in the Fig. 5 of Ref.~\citenum{Gol18} where the full frequency and standard contour-deformation approaches were compared in the study of $H_2O$ 1\textit{s} core state.  Our def2-QZVP $G_0W_0$@PBE quasiparticle energy at $-523.80\,eV$ ($Z_{QP}=0.20$) is identical to the   $-523.8\,eV$ def2-QZVP $G_0W_0$@PBE value of Ref.~\citenum{Gol18}.

As expected, the analytic continuation based on sampling data points along the imaginary-axis only (pink dotted-dashed lines) fails in reproducing faithfully this complex self-energy structure. Still, we observe that with our limited n$\omega$=14 sampling of the imaginary axis, the resulting AC on $W$ leads to a few structures in the vicinity of the QP peak. This contrasts with the results obtained with the AC of $\Sigma^C$   (Ref. \citenum{Gov18}) that does not capture any pole in this energy range, even with a very large (1024 pts) sampling set along the imaginary axis.

We did not attempt here to optimize the sampling strategy, by e.g. differentiating sampling grids and $\Delta \omega$ spacing associated with valence and core residues. However, Fig.~\ref{water_core} demonstrates that Fig.~\ref{fig_W_sampling}b sampling strategy with a coarse $\Delta \omega$ sampling of $1\,eV$ already provides a well converged spectral function at moderate cost. 

\section{Conclusion}

 The contour-deformation approach to calculating the $GW$ quasiparticle energies requires the knowledge of the screened Coulomb potential $W$ along the imaginary frequency axis but also at specific energies on the real-frequency axis. For these latter contributions, we have shown that the analytic continuation from the imaginary to the real-frequency axis of the screened Coulomb potential $W(\omega)$ leads to a much more robust scheme as compared to current techniques involving the direct continuation of the  self-energy $\Sigma(E)$. As a result, the present approach  allows to reach the meV accuracy for the calculation of the HOMO and LUMO levels energy of the $GW$100 molecular set, \cite{Set15} calculating only a very few ($n\omega \simeq 14$)  $W(i\omega)$ matrices along the imaginary axis.  In particular, all difficulties previously observed for selected systems (MgO, BN, O$_3$, BeO) are now resolved. Even better results are obtained with a  recent set of larger medium-size real-life acceptor molecules. \cite{Kni16} In this latter case, the error associated  with the analytic continuation from the imaginary axis is  reduced to about $10^{-2}$ meV for frontier orbitals when starting with PBE0 input eigenstates. 

The calculation of the $GW$ correction to states located far away from the gap remains a challenging issue. The difficulties originate  from the expected behaviour of such quasiparticle states with the decay of the main quasiparticle peak over a very dense  forest  of poles along the real axis, yielding a very large number of solutions of the quasiparticle equation associated with very small $Z_{QP}$ renormalization  factors. The number and location of such peaks are unphysical, being related in particular to the finiteness of the Kohn-Sham basis set adopted to generate the input one-body eigenstates. However, as expected, these peaks merge  rapidly into broad quasiparticle structures if one calculates the spectral function $A_n^{GW}(\omega)$  along $\omega$-lines parallel to the real-axis but with a finite $\Im(\omega)$ imaginary part. 

For such broad structures, the analytic continuation of $W$ from the imaginary-axis to the $\Im(\omega)\simeq100$ meV axis yields errors below 10 meV for states within typically an energy gap of frontier orbitals, but the error can become as large as an eV further away in energy. In such situations, the analytic continuation scheme can be consolidated by calculating a few screened Coulomb potential matrices in the complex plane, increasing the accuracy of the analytic continuation in the vicinity of each added reference point. In practice, adding  a very coarse grid ($\Delta \omega \simeq 1$ eV) of reference  $W(\omega)$ calculated  along an  $\omega$-line parallel to the real axis, allows to bring the error of the present analytic continuation approach below 10 meV over a much larger energy range. Contrary to the self-energy operator, we emphasize that calculating $W(\omega)$ at frequencies anywhere in the complex plane bears the same computational cost. The same strategy can be used for core states as demonstrated in the paradigmatic case of the H$_2$O 1\textit{s} level. Sampling separately the energy ranges spanned by core and valence poles allows for an efficient yet accurate calculations of the core state spectral function  over a large energy window. 

Difficulties remain  occasionally with states far away from the gap for which there are no clear quasiparticle peaks with sizeable $Z_{QP}$ weight dominating the spectral function.  In such situations, the analytic continuation of $W$ using a coarse grid of data points in the complex plane remains accurate, but instabilities can occur when attempting to extract automatically one peak out of several equivalent structures. As such, the analytic continuation scheme is not to be blamed, but rather the concept of quasiparticles. Such cases question the use of self-consistent approaches where the quasiparticle energy of low-lying states is re-injected to build an updated Green's function $G$ and screened potential $W$. These   difficulties are well beyond the scope of the present paper.

\begin{suppinfo}

 We provide in the Supporting Information the calculated ionization potential and electronic affinity for the $GW$100 test set and the 24 medium size acceptor molecular set introduced in Ref.~\citenum{Kni16}. We further provide the AC errors as in Fig.~\ref{QP_error} but on a larger energy range that contains all occupied and unoccupied states.

\end{suppinfo}

\bibliography{manuscript}

\begin{mcitethebibliography}{85}
\bibitem[cen()]{centersolution}
We do not consider the "center" values of Ref.~\citenum{Set15} that correspond
  to the unphysical portions of $\Sigma$ connecting the $\pm \infty$ values on
  each side of the poles of the $GW$ operator.\relax
\mciteBstWouldAddEndPunctfalse
\mciteSetBstMidEndSepPunct{\mcitedefaultmidpunct}
{}{\mcitedefaultseppunct}\relax
\EndOfBibitem
\bibitem[ana()]{analytic0}
The standard technique relying on performing the analytic continuation of the
  self-energy may also be improved by calculating reference $\Sigma(\omega)$
  self-energies away from the imaginary-axis. We re-emphasize that calculating
  $W$ always scales as $\mathcal{O}(N^4)$ for any frequency in the complex
  plane, while calculating the self-energy operator scales as
  $\mathcal{O}(N^5)$ for states away from the gap using the standard
  contour-deformation approach, and $\mathcal{O}(N^6)$ if the exact spectral
  representation of $W$ is obtained by full diagonalization of the
  susceptibility.\relax
\mciteBstWouldAddEndPunctfalse
\mciteSetBstMidEndSepPunct{\mcitedefaultmidpunct}
{}{\mcitedefaultseppunct}\relax
\EndOfBibitem
\end{mcitethebibliography}

\begin{tocentry}
\includegraphics[width=\linewidth]{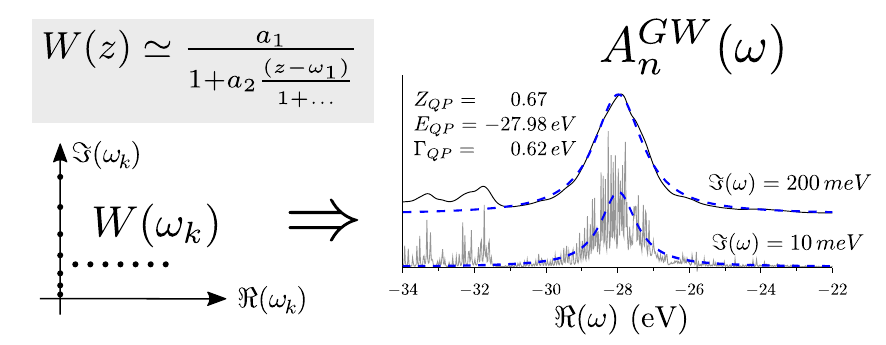}   
\vskip .2cm  
The analytic continuation of the screened Coulomb potential from few selected data points yields accurate $GW$ self-energies over large energy ranges. 
\end{tocentry}

\end{document}



We provide here below the ionization potentials (Section~\ref{IPGW100}) and electronic affinities (Section~\ref{AEGW100}) for the GW$100$ test set, \cite{Set15}  comparing reference calculations (Fig.~2b scheme, $\Delta \omega=125 meV$) with data stemming from the analytic continuation of $W$ relying on a reduced n$\omega$=14 data points sampling along the imaginary axis (Fig.~2a scheme). We consider only the 93 molecules not requiring pseudopotentials.  Both def2-QZVP $G_0W_0$@PBE and def2-QZVP $G_0W_0$@PBE0 data are presented. Similarly, we list the aug-cc-pVTZ $G_0W_0$@PBE0 ionization potential and electronic affinities (Section~\ref{IPMarom}) for the set of 24 medium size acceptor molecules from Refs.~\citenum{Kni16,She16}. Finally, we reproduce Fig.~7b of the main manuscript, but over the full energy range and a lower $Z=0.2$ cutoff value.


\section{$GW$100 test set ionisation potentials } 
\label{IPGW100}

\begin{spacing}{1}
\begin{longtable}[h]{rlrrrr}

 & & \multicolumn{4}{c}{GW100 IPs}\\
 \hline
 & & \multicolumn{2}{c}{$G_0W_0$@PBE} & \multicolumn{2}{c}{$G_0W_0$@PBE0} \\
 i & Formula & ref. & P-14 & ref. & P-14 \\
 \hline
 \endfirsthead

 \multicolumn{6}{c}{Continuation of GW100 IPs}\\
 \hline
 & &  \multicolumn{2}{c}{$G_0W_0$@PBE} & \multicolumn{2}{c}{$G_0W_0$@PBE0} \\
 i & Formula & ref. & P-14 & ref. & P-14 \\
 \hline
 \endhead

 1     &    He                           &    23.4\hl{77}   &       23.4\hl{83}   &   23.99\hl{5} &    23.99\hl{9} \\  
 2     &    Ne                           &    20.376        &       20.376        &   20.879      &    20.879      \\  
 3     &    Ar                           &    15.132        &       15.132        &   15.417      &    15.417      \\  
 4     &    Kr                           &    13.571        &       13.571        &   13.781      &    13.781      \\  
 6     &    H2                           &    15.815        &       15.815        &   16.182      &    16.182      \\  
 7     &    Li$_{2}$                     &     4.992        &        4.992        &    5.247      &     5.247      \\  
 8     &    Na$_{2}$                     &     4.837        &        4.837        &    4.966      &     4.966      \\  
 9     &    Na$_{4}$                     &     4.120        &        4.120        &    4.233      &     4.233      \\  
 10    &    Na$_{6}$                     &     4.242        &        4.242        &    4.356      &     4.356      \\  
 11    &    K$_{2}$                      &     3.98\hl{0}   &        3.98\hl{1}   &    4.072      &     4.072      \\  
 13    &    N$_{2}$                      &    14.889        &       14.889        &   15.429      &    15.429      \\  
 14    &    P$_{2}$                      &    10.205        &       10.205        &   10.373      &    10.373      \\  
 15    &    As$_{2}$                     &     9.471        &        9.471        &    9.599      &     9.599      \\  
 16    &    F$_{2}$                      &    14.962        &       14.962        &   15.529      &    15.529      \\  
 17    &    Cl$_{2}$                     &    11.101        &       11.101        &   11.384      &    11.384      \\  
 18    &    Br$_{2}$                     &    10.217        &       10.217        &   10.451      &    10.451      \\  
 20    &    CH$_{4}$                     &    13.927        &       13.927        &   14.267      &    14.267      \\  
 21    &    C$_{2}$H$_{6}$               &    12.365        &       12.365        &   12.676      &    12.676      \\  
 22    &    C$_{3}$H$_{8}$               &    11.795        &       11.795        &   12.106      &    12.106      \\  
 23    &    C$_{4}$H$_{10}$              &    11.490        &       11.490        &   11.804      &    11.804      \\  
 24    &    C$_{2}$H$_{4}$               &    10.325        &       10.325        &   10.514      &    10.514      \\  
 25    &    C$_{2}$H$_{2}$               &    11.020        &       11.020        &   11.271      &    11.271      \\  
 26    &    C$_{4}$                      &    10.781        &       10.781        &   11.156      &    11.156      \\  
 27    &    C$_{3}$H$_{6}$               &    10.555        &       10.555        &   10.821      &    10.821      \\  
 28    &    C$_{6}$H$_{6}$               &     8.985        &        8.985        &    9.196      &     9.196      \\  
 29    &    C$_{8}$H$_{8}$               &     8.055        &        8.055        &    8.284      &     8.284      \\  
 30    &    C$_{5}$H$_{6}$               &     8.349        &        8.349        &    8.551      &     8.551      \\  
 31    &    C$_{2}$H$_{3}$F              &    10.197        &       10.197        &   10.441      &    10.441      \\  
 32    &    C$_{2}$H$_{3}$Cl             &     9.761        &        9.761        &   10.008      &    10.008      \\  
 33    &    C$_{2}$H$_{3}$Br             &     8.990        &        8.990        &    9.184      &     9.184      \\  
 33    &    C$_{2}$H$_{3}$Br             &     9.491        &        9.491        &    9.728      &     9.728      \\  
 35    &    CF$_{4}$                     &    15.366        &       15.366        &   15.995      &    15.995      \\  
 36    &    CCl$_{4}$                    &    10.975        &       10.975        &   11.401      &    11.401      \\  
 37    &    CBr$_{4}$                    &     9.899        &        9.899        &   10.272      &    10.272      \\  
 39    &    SiH$_{4}$                    &    12.310        &       12.310        &   12.699      &    12.699      \\  
 40    &    GeH$_{4}$                    &    12.022        &       12.022        &   12.358      &    12.358      \\  
 41    &    Si$_{2}$H$_{6}$              &    10.309        &       10.309        &   10.585      &    10.585      \\  
 42    &    Si$_{5}$H$_{12}$             &     8.934        &        8.934        &    9.220      &     9.220      \\  
 43    &    LiH                          &     6.552        &        6.552        &    7.547      &     7.547      \\  
 44    &    KH                           &     4.862        &        4.862        &    5.660      &     5.660      \\  
 45    &    BH$_{3}$                     &    12.867        &       12.867        &   13.225      &    13.225      \\  
 46    &    B$_{2}$H$_{6}$               &    11.835        &       11.835        &   12.218      &    12.218      \\  
 47    &    NH$_{3}$                     &    10.315        &       10.315        &   10.698      &    10.698      \\  
 48    &    HN$_{3}$                     &    10.395        &       10.395        &   10.686      &    10.686      \\  
 49    &    PH$_{3}$                     &    10.273        &       10.273        &   10.487      &    10.487      \\  
 50    &    AsH$_{3}$                    &    10.123        &       10.123        &   10.300      &    10.300      \\  
 51    &    SH$_{2}$                     &    10.031        &       10.031        &   10.248      &    10.248      \\  
 52    &    FH                           &    15.302        &       15.302        &   15.748      &    15.748      \\  
 53    &    ClH                          &    12.246        &       12.246        &   12.501      &    12.501      \\  
 54    &    LiF                          &     9.94\hl{6}   &        9.94\hl{7}   &   10.829      &    10.829      \\  
 55    &    F$_{2}$Mg                    &    12.323        &       12.323        &   13.188      &    13.188      \\  
 56    &    TiF$_{4}$                    &    13.896        &       13.896        &   14.868      &    14.868      \\  
 57    &    AlF$_{3}$                    &    14.252        &       14.252        &   14.950      &    14.950      \\  
 58    &    BF                           &    10.562        &       10.562        &   10.909      &    10.909      \\  
 59    &    SF$_{4}$                     &    12.117        &       12.117        &   12.558      &    12.558      \\  
 60    &    BrK                          &     7.30\hl{4}   &        7.30\hl{5}   &    7.946      &     7.946      \\  
 61    &    GaCl                         &     9.554        &        9.554        &    9.719      &     9.719      \\  
 62    &    NaCl                         &     8.09\hl{8}   &        8.09\hl{9}   &    8.835      &     8.835      \\  
 63    &    MgCl$_{2}$                   &    10.991        &       10.991        &   11.462      &    11.462      \\  
 65    &    BN                           &    11.010        &       11.010        &   11.543      &    11.543      \\  
 66    &    NCH                          &    13.211        &       13.211        &   13.525      &    13.525      \\  
 67    &    PN                           &    11.136        &       11.136        &   11.688      &    11.688      \\  
 68    &    H$_{2}$NNH$_{2}$             &     9.281        &        9.281        &    9.647      &     9.647      \\  
 69    &    H$_{2}$CO                    &    10.328        &       10.328        &   10.784      &    10.784      \\  
 70    &    CH$_{4}$O                    &    10.562        &       10.562        &   10.974      &    10.974      \\  
 71    &    C$_{2}$H$_{6}$O              &    10.157        &       10.157        &   10.604      &    10.604      \\  
 72    &    C$_{2}$H$_{4}$O              &     9.548        &        9.548        &   10.094      &    10.094      \\  
 73    &    C$_{4}$H$_{10}$O             &     9.319        &        9.319        &    9.777      &     9.777      \\  
 74    &    CH$_{2}$O$_{2}$              &    10.730        &       10.730        &   11.292      &    11.292      \\  
 75    &    HOOH                         &    10.987        &       10.987        &   11.444      &    11.444      \\  
 76    &    H$_{2}$O                     &    11.973        &       11.973        &   12.387      &    12.387      \\  
 77    &    CO$_{2}$                     &    13.250        &       13.250        &   13.662      &    13.662      \\  
 78    &    CS$_{2}$                     &     9.748        &        9.748        &    9.984      &     9.984      \\  
 79    &    CSO                          &    10.906        &       10.906        &   11.170      &    11.170      \\  
 80    &    CSeO                         &    10.199        &       10.199        &   10.413      &    10.413      \\  
 81    &    CO                           &    13.571        &       13.571        &   14.120      &    14.120      \\  
 82    &    O$_{3}$                      &    11.967        &       11.967        &   12.563      &    12.563      \\  
 83    &    SO$_{2}$                     &    11.823        &       11.823        &   12.272      &    12.272      \\  
 84    &    BeO                          &     9.63\hl{3}   &        9.63\hl{5}   &    9.644      &     9.644      \\  
 85    &    MgO                          &     6.671        &        6.671        &    7.457      &     7.457      \\  
 86    &    C$_{7}$H$_{8}$               &     8.612        &        8.612        &    8.818      &     8.818      \\  
 87    &    C$_{8}$H$_{10}$              &     8.553        &        8.553        &    8.766      &     8.766      \\  
 88    &    C$_{6}$F$_{6}$               &     9.493        &        9.493        &    9.893      &     9.893      \\  
 89    &    C$_{6}$H$_{5}$OH             &     8.366        &        8.366        &    8.609      &     8.609      \\  
 89    &    C$_{6}$H$_{5}$OH             &     8.263        &        8.263        &    8.502      &     8.502      \\  
 90    &    C$_{6}$H$_{5}$NH$_{2}$       &     7.644        &        7.644        &    7.914      &     7.914      \\  
 91    &    C$_{5}$H$_{5}$N              &     9.040        &        9.040        &    9.600      &     9.600      \\  
 92    &    C$_{5}$H$_{5}$N$_{5}$O       &     7.691        &        7.691        &    7.976      &     7.976      \\  
 93    &    C$_{5}$H$_{5}$N$_{5}$O       &     7.975        &        7.975        &    8.256      &     8.256      \\  
 94    &    C$_{4}$H$_{5}$N$_{3}$O       &     8.287        &        8.287        &    8.683      &     8.683      \\  
 95    &    C$_{5}$H$_{6}$N$_{2}$O$_{2}$ &     8.707        &        8.707        &    9.057      &     9.057      \\  
 96    &    C$_{4}$H$_{4}$N$_{2}$O$_{2}$ &     9.223        &        9.223        &    9.457      &     9.457      \\  
 97    &    CH$_{4}$N$_{2}$O             &     9.321        &        9.321        &    9.908      &     9.908      \\  
 99    &    Cu$_{2}$                     &     7.529        &        7.529        &    7.551      &     7.551      \\  
100    &    CuCN                         &     9.42\hl{3}   &        9.42\hl{5}   &   10.253      &    10.253      \\

\hline
\caption{Ionization potential (defined as -$E_{HOMO}$) for the $GW$100 test set as obtained with the present analytic continuation scheme. Reference calculations (ref.) are associated with the fine ($\Delta \omega=125\,\mathrm{m}eV$)  Fig.~2b sampling scheme. The  (P-14) calculations use only the $n\omega$=14 reference $W(i\omega)$ along the imaginary axis from Fig.~2a sampling scheme. Calculations are performed at the def2-QZVP level with Coulomb fitting resolution-of-the-identity using the corresponding def2-QZVP-RI auxiliary basis. Both PBE and PBE0 Kohn-Sham DFT starting points are presented. Digits differing between the (ref.) and (P-14) data are highlighted in red. }
\label{tab:GW100_IP_details}
\end{longtable}
\end{spacing}

\section{$GW$100 test set electronic affinities } 
\label{AEGW100}

\begin{spacing}{1}
\begin{longtable}[h]{rlrrrr}

 & & \multicolumn{4}{c}{GW100 EAs}\\
 \hline
 & &  \multicolumn{2}{c}{$G_0W_0$@PBE} & \multicolumn{2}{c}{$G_0W_0$@PBE0} \\
 i & Formula & ref. & P-14 & ref. & P-14 \\
 \hline
 \endfirsthead

 \multicolumn{6}{c}{Continuation of GW100 EAs}\\
 \hline
 & &  \multicolumn{2}{c}{$G_0W_0$@PBE} & \multicolumn{2}{c}{$G_0W_0$@PBE0} \\
 i & Formula & ref. & P-14 & ref. & P-14 \\
 \hline
 \endhead

 1     &    He                           & -11.00\hl{6} & -11.00\hl{7} &  -11.01\hl{9} &  -11.02\hl{0} \\ 
 2     &    Ne                           &  -11.643     &  -11.643     &  -11.674      &  -11.674      \\ 
 3     &    Ar                           &   -8.113     &   -8.113     &   -8.166      &   -8.166      \\ 
 4     &    Kr                           &   -7.635     &   -7.635     &   -7.689      &   -7.689      \\ 
 6     &    H2                           &   -3.504     &   -3.504     &   -3.436      &   -3.436      \\ 
 7     &    Li$_{2}$                     &    0.626     &    0.626     &    0.409      &    0.409      \\ 
 8     &    Na$_{2}$                     &    0.550     &    0.550     &    0.421      &    0.421      \\ 
 9     &    Na$_{4}$                     &    1.009     &    1.009     &    0.779      &    0.779      \\ 
 10    &    Na$_{6}$                     &    0.971     &    0.971     &    0.762      &    0.762      \\ 
 11    &    K$_{2}$                      &    0.647     &    0.647     &    0.508      &    0.508      \\ 
 13    &    N$_{2}$                      &   -2.449     &   -2.449     &   -2.586      &   -2.586      \\ 
 14    &    P$_{2}$                      &    0.724     &    0.724     &    0.608      &    0.608      \\ 
 15    &    As$_{2}$                     &    0.847     &    0.847     &    0.753      &    0.753      \\ 
 16    &    F$_{2}$                      &    0.704     &    0.704     &    0.308      &    0.308      \\ 
 17    &    Cl$_{2}$                     &    0.887     &    0.887     &    0.687      &    0.687      \\ 
 18    &    Br$_{2}$                     &    1.396     &    1.396     &    1.224      &    1.224      \\ 
 20    &    CH$_{4}$                     &   -2.450     &   -2.450     &   -2.486      &   -2.486      \\ 
 21    &    C$_{2}$H$_{6}$               &   -2.294     &   -2.294     &   -2.357      &   -2.357      \\ 
 22    &    C$_{3}$H$_{8}$               &   -2.195     &   -2.195     &   -2.278      &   -2.278      \\ 
 23    &    C$_{4}$H$_{10}$              &   -2.145     &   -2.145     &   -2.244      &   -2.244      \\ 
 24    &    C$_{2}$H$_{4}$               &   -2.020     &   -2.020     &   -2.189      &   -2.189      \\ 
 25    &    C$_{2}$H$_{2}$               &   -2.863     &   -2.863     &   -2.965      &   -2.965      \\ 
 26    &    C$_{4}$                      &    2.936     &    2.936     &    2.710      &    2.710      \\ 
 27    &    C$_{3}$H$_{6}$               &   -2.453     &   -2.453     &   -2.525      &   -2.525      \\ 
 28    &    C$_{6}$H$_{6}$               &   -1.087     &   -1.087     &   -1.300      &   -1.300      \\ 
 29    &    C$_{8}$H$_{8}$               &   -0.059     &   -0.059     &   -0.316      &   -0.316      \\ 
 30    &    C$_{5}$H$_{6}$               &   -1.040     &   -1.040     &   -1.274      &   -1.274      \\ 
 31    &    C$_{2}$H$_{3}$F              &   -2.146     &   -2.146     &   -2.317      &   -2.317      \\ 
 32    &    C$_{2}$H$_{3}$Cl             &   -1.425     &   -1.425     &   -1.623      &   -1.623      \\ 
 33    &    C$_{2}$H$_{3}$Br             &   -1.380     &   -1.380     &   -1.571      &   -1.571      \\ 
 33    &    C$_{2}$H$_{3}$Br             &   -1.257     &   -1.257     &   -1.472      &   -1.472      \\ 
 35    &    CF$_{4}$                     &   -4.413     &   -4.413     &   -4.479      &   -4.479      \\ 
 36    &    CCl$_{4}$                    &    0.008     &    0.008     &   -0.245      &   -0.245      \\ 
 37    &    CBr$_{4}$                    &    1.079     &    1.079     &    0.856      &    0.856      \\ 
 39    &    SiH$_{4}$                    &   -2.508     &   -2.508     &   -2.560      &   -2.560      \\ 
 40    &    GeH$_{4}$                    &   -2.300     &   -2.300     &   -2.391      &   -2.391      \\ 
 41    &    Si$_{2}$H$_{6}$              &   -1.686     &   -1.686     &   -1.889      &   -1.889      \\ 
 42    &    Si$_{5}$H$_{12}$             &   -0.161     &   -0.161     &   -0.416      &   -0.416      \\ 
 43    &    LiH                          &    0.072     &    0.072     &    0.092      &    0.092      \\ 
 44    &    KH                           &    0.180     &    0.180     &    0.154      &    0.154      \\ 
 45    &    BH$_{3}$                     &   -0.115     &   -0.115     &   -0.263      &   -0.263      \\ 
 46    &    B$_{2}$H$_{6}$               &   -0.838     &   -0.838     &   -1.014      &   -1.014      \\ 
 47    &    NH$_{3}$                     &   -2.312     &   -2.312     &   -2.345      &   -2.345      \\ 
 48    &    HN$_{3}$                     &   -1.399     &   -1.399     &   -1.549      &   -1.549      \\ 
 49    &    PH$_{3}$                     &   -2.496     &   -2.496     &   -2.096      &   -2.096      \\ 
 50    &    AsH$_{3}$                    &   -2.320     &   -2.320     &   -2.015      &   -2.015      \\ 
 51    &    SH$_{2}$                     &   -2.557     &   -2.557     &   -2.024      &   -2.024      \\ 
 52    &    FH                           &   -2.543     &   -2.543     &   -2.483      &   -2.483      \\ 
 53    &    ClH                          &   -2.064     &   -2.064     &   -2.051      &   -2.051      \\ 
 54    &    LiF                          &   -0.091     &   -0.091     &    0.049      &    0.049      \\ 
 55    &    F$_{2}$Mg                    &    0.138     &    0.138     &    0.084      &    0.084      \\ 
 56    &    TiF$_{4}$                    &    0.598     &    0.598     &    0.675      &    0.675      \\ 
 57    &    AlF$_{3}$                    &   -0.157     &   -0.157     &   -0.271      &   -0.271      \\ 
 58    &    BF                           &   -1.216     &   -1.216     &   -1.298      &   -1.298      \\ 
 59    &    SF$_{4}$                     &   -0.377     &   -0.377     &   -0.543      &   -0.543      \\ 
 60    &    BrK                          &    0.307     &    0.307     &    0.362      &    0.362      \\ 
 61    &    GaCl                         &    0.020     &    0.020     &   -0.062      &   -0.062      \\ 
 62    &    NaCl                         &    0.394     &    0.394     &    0.433      &    0.433      \\ 
 63    &    MgCl$_{2}$                   &    0.430     &    0.430     &    0.333      &    0.333      \\ 
 65    &    BN                           &    3.947     &    3.947     &    3.760      &    3.760      \\ 
 66    &    NCH                          &   -2.577     &   -2.577     &   -2.674      &   -2.674      \\ 
 67    &    PN                           &    0.202     &    0.202     &    0.084      &    0.084      \\ 
 68    &    H$_{2}$NNH$_{2}$             &   -1.993     &   -1.993     &   -2.062      &   -2.062      \\ 
 69    &    H$_{2}$CO                    &   -0.959     &   -0.959     &   -1.192      &   -1.192      \\ 
 70    &    CH$_{4}$O                    &   -2.247     &   -2.247     &   -2.264      &   -2.264      \\ 
 71    &    C$_{2}$H$_{6}$O              &   -2.076     &   -2.076     &   -2.130      &   -2.130      \\ 
 72    &    C$_{2}$H$_{4}$O              &   -1.050     &   -1.050     &   -1.375      &   -1.375      \\ 
 73    &    C$_{4}$H$_{10}$O             &   -2.100     &   -2.100     &   -2.199      &   -2.199      \\ 
 74    &    CH$_{2}$O$_{2}$              &   -1.910     &   -1.910     &   -2.172      &   -2.172      \\ 
 75    &    HOOH                         &   -2.349     &   -2.349     &   -2.548      &   -2.548      \\ 
 76    &    H$_{2}$O                     &   -2.370     &   -2.370     &   -2.377      &   -2.377      \\ 
 77    &    CO$_{2}$                     &   -2.497     &   -2.497     &   -2.509      &   -2.509      \\ 
 78    &    CS$_{2}$                     &    0.197     &    0.197     &    0.104      &    0.104      \\ 
 79    &    CSO                          &   -1.215     &   -1.215     &   -1.360      &   -1.360      \\ 
 80    &    CSeO                         &   -0.868     &   -0.868     &   -1.024      &   -1.024      \\ 
 81    &    CO                           &   -0.671     &   -0.671     &   -0.797      &   -0.797      \\ 
 82    &    O$_{3}$                      &    2.296     &    2.296     &    2.212      &    2.212      \\ 
 83    &    SO$_{2}$                     &    1.002     &    1.002     &    0.883      &    0.883      \\ 
 84    &    BeO                          &    2.493     &    2.493     &    2.161      &    2.161      \\ 
 85    &    MgO                          &    1.893     &    1.893     &    1.737      &    1.737      \\ 
 86    &    C$_{7}$H$_{8}$               &   -1.016     &   -1.016     &   -1.237      &   -1.237      \\ 
 87    &    C$_{8}$H$_{10}$              &   -1.044     &   -1.044     &   -1.279      &   -1.279      \\ 
 88    &    C$_{6}$F$_{6}$               &   -0.657     &   -0.657     &   -0.633      &   -0.633      \\ 
 89    &    C$_{6}$H$_{5}$OH             &   -0.959     &   -0.959     &   -1.155      &   -1.155      \\ 
 89    &    C$_{6}$H$_{5}$OH             &   -1.013     &   -1.013     &   -1.211      &   -1.211      \\ 
 90    &    C$_{6}$H$_{5}$NH$_{2}$       &   -1.148     &   -1.148     &   -1.358      &   -1.358      \\ 
 91    &    C$_{5}$H$_{5}$N              &   -0.514     &   -0.514     &   -0.735      &   -0.735      \\ 
 92    &    C$_{5}$H$_{5}$N$_{5}$O       &   -0.749     &   -0.749     &   -1.008      &   -1.008      \\ 
 93    &    C$_{5}$H$_{5}$N$_{5}$O       &   -0.474     &   -0.474     &   -0.735      &   -0.735      \\ 
 94    &    C$_{4}$H$_{5}$N$_{3}$O       &   -0.263     &   -0.263     &   -0.476      &   -0.476      \\ 
 95    &    C$_{5}$H$_{6}$N$_{2}$O$_{2}$ &   -0.057     &   -0.057     &   -0.282      &   -0.282      \\ 
 96    &    C$_{4}$H$_{4}$N$_{2}$O$_{2}$ &   -0.012     &   -0.012     &   -0.229      &   -0.229      \\ 
 97    &    CH$_{4}$N$_{2}$O             &   -1.621     &   -1.621     &   -1.702      &   -1.702      \\ 
 99    &    Cu$_{2}$                     &    0.756     &    0.756     &    0.595      &    0.595      \\ 
100    &    CuCN                         &    1.647     &    1.647     &    1.264      &    1.264      \\ 
\hline

\caption{Same as in Table S1 but for the electronic affinities (defined as -$E_{LUMO}$).  }
\label{tab:GW100_EA_details}
\end{longtable}
\end{spacing}

\section{ Acceptor molecular set (Ref.~\citenum{Kni16} ) ionization potential and electronic affinity  } 
\label{IPMarom}

\begin{table}[h!]
\begin{tabular}{lrrrrrr}
& \multicolumn{2}{c}{CCSD(T)} & \multicolumn{2}{c}{$G_0W_0$}  & \multicolumn{2}{c}{eg$GW$} \\ 
& IP  &  EA & IP & EA & IP & EA \\
\hline
Anthracene                         &     7.474    &    0.260    &     7.156    &     0.561    &      7.299    &     0.372   \\
Acridine                           &       N/A    &      N/A    &     7.637    &     0.897    &      7.811    &     0.715   \\
Phenazine                          &     8.417    &    1.030    &     8.052    &     1.317    &      8.248    &     1.145   \\
Azulene                            &     7.485    &    0.482    &     7.203    &     0.717    &      7.355    &     0.538   \\
Benzoquinone                       &    10.172    &    1.458    &     9.811    &     1.746    &     10.409    &     1.539   \\
Naphthalenedione                   &     9.791    &    1.388    &     9.405    &     1.681    &     10.007    &     1.445   \\
Dichlone                           &     9.885    &    1.819    & {\bf 9.480}  &     2.117    &      9.724    &     1.896   \\
F4-benzoquinone            &    11.038    &    2.177    &    10.677    &     2.479    &     10.945    &     2.291   \\
Cl4-benzoquinone            &    10.122    &    2.360    &     9.793    &     2.666    &     10.031    &     2.457   \\
Nitrobenzene                       &    10.139    &    0.442    &     9.851    &     0.748    &     10.064    &     0.479   \\
F4-benzenedicar.   &    10.664    &    1.513    &    10.316    &     1.843    &     10.568    &     1.667   \\
Dinitrobenzoni.                &    11.074    &    1.666    &    10.812    &     2.023    &     11.062    &     1.753   \\
Nitrobenzoni.                  &    10.548    &    1.200    &    10.262    &     1.560    &     10.492    &     1.320   \\
Benzonitrile                       &     9.881    &   -0.287    &     9.573    &    -0.020    &      9.783    &    -0.237   \\
Fumaronitrile                      &    11.399    &    0.892    &    11.026    &     1.208    &     11.282    &     0.984   \\
mDCNB                              &    10.372    &    0.540    &    10.077    &     0.857    &     10.303    &     0.631   \\
TCNE                               &    11.898    &    2.944    &    11.518    &     3.344    &     11.774    &     3.173   \\
TCNQ                               &     9.492    &    3.230    &     9.199    & {\bf 3.669}  &      9.345    & {\bf 3.558} \\
Maleicanhydride                    &    11.226    &    0.918    &    10.851    &     1.150    &     11.445    &     0.918   \\
Phthalimide                        &    10.023    &    0.541    &     9.747    &     0.787    & {\bf 10.358}  &     0.562   \\
phthalicanhydride                  &    10.477    &    0.777    &    10.178    &     1.043    &     10.388    &     0.810   \\
Cl4-isobenz.    &       N/A    &      N/A    &     9.651    &     1.843    &      9.868    &     1.614   \\
NDCA                               &       N/A    &      N/A    &     8.775    &     1.460    &      8.938    &     1.265   \\
bodipy                             &       N/A    &      N/A    &     7.875    &     1.758    &      7.997    &     1.612   \\
\hline
MaxErr   &      &      &  -0.405  &  0.439  & 0.335 & 0.328 \\
MAE      &       &      &  0.329  & 0.307   & 0.135 & 0.098 \\
MSE      &       &      &  -0.329 & 0.308 & -0.035 & 0.098 \\
\hline
\end{tabular}
\caption{Ionization potentials (IP) and electronic affinities (EA) for the 24 molecules of  Refs.~\citenum{Kni16,She16}. Reference CCSD(T) calculations are from Ref.~\cite{She16}. 
All calculations are performed at the aug-cc-pVTZ level. $GW$ calculations provided are based on PBE0 Kohn-Sham eigenstates, following the "reference" calculations scheme of Fig.~2B main text with $\Delta \omega$=125 meV. The eg$GW$ data indicate gap-self-consistent (scissor-like) calculations. The IP and AE yielding the largest discrepancy (Max.Err.) with the CCSD(T) data are highlighted.}
\label{tab:marom_set_details_SI}
\end{table}

\section{Analytic continuation error}

\begin{figure}[h!]
\includegraphics[width=8cm]{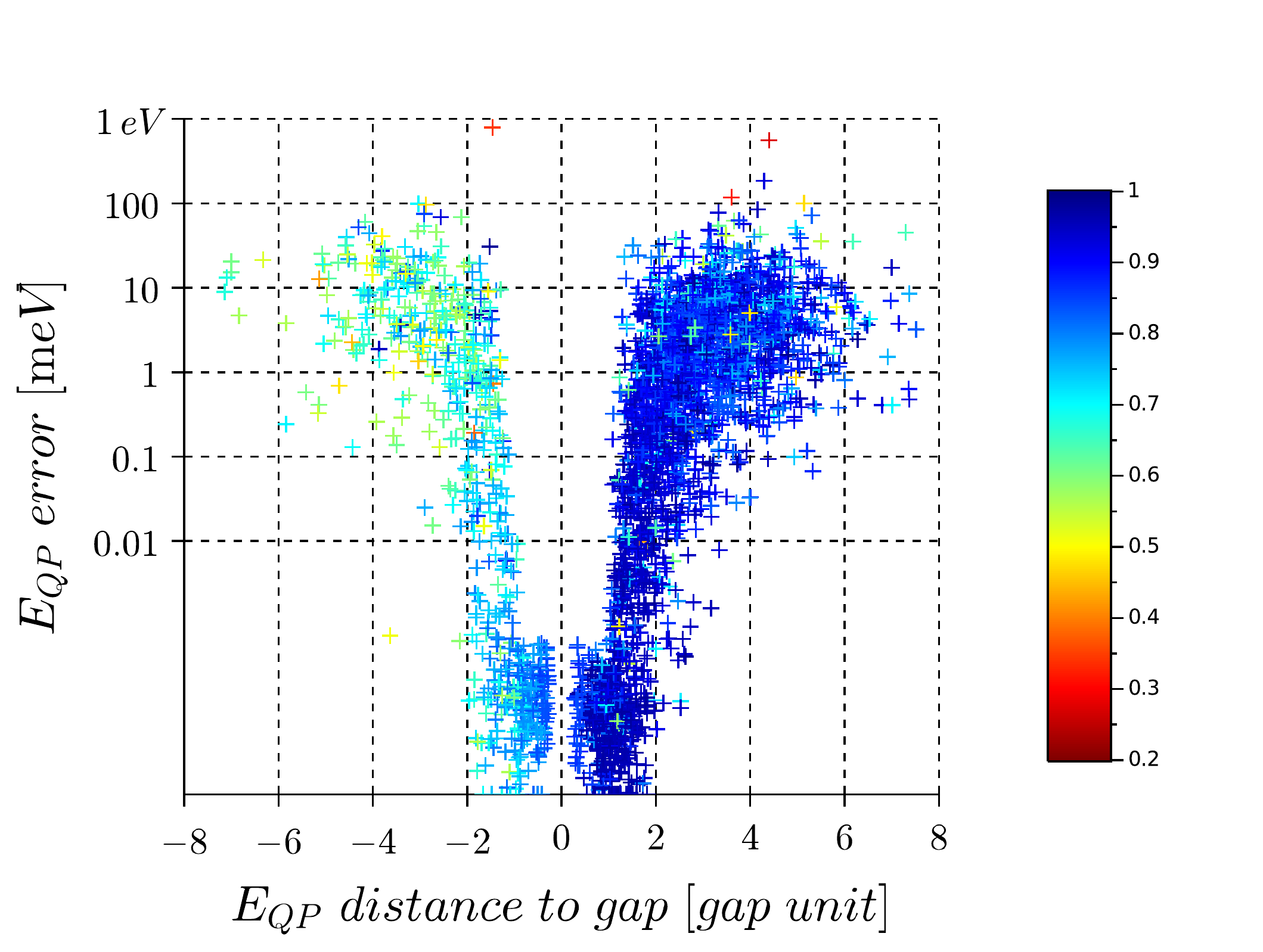}
\caption{Same as Fig.~7b of the main manuscript, but over the full energy range and with a lower $Z=0.2$ cutoff value. For occupied states, no new outliers are observed. In the unoccupied manifold, the main outlier corresponds to the LUMO+139 of acridine located 16eV above the LUMO $G_0W_0$@PBE0 value.}
\label{fig_W_sampling}
\end{figure}

\bibliography{manuscript}